\documentclass[twocolumn]{aastex631}
\usepackage{CJK}

\let\tablenum\relax
\usepackage{siunitx}
\usepackage{booktabs}

\usepackage{xcolor, soul}

\shorttitle{web-feeding model in the COSMOS2020}
\shortauthors{Ko et al.}

\begin{document}

\begin{CJK*}{UTF8}{mj}

\title{Test of  Cosmic Web-Feeding Model for Star Formation in Galaxy Clusters in the COSMOS Field}
\author{Eunhee Ko (고은희)}
\affiliation{SNU Astronomy Research Center, Astronomy Program, Department of Physics and Astronomy, Seoul National University, Seoul 08826, Republic of Korea}
\affiliation{Institut d’Astrophysique de Paris, UMR 7095, CNRS, Sorbonne Universit\'e, 98 bis boulevard Arago, F-75014 Paris, France}

\author{Myungshin Im}
\affiliation{SNU Astronomy Research Center, Astronomy Program, Department of Physics and Astronomy, Seoul National University, Seoul 08826, Republic of Korea}

\author{Seong-kook Lee}
\affiliation{SNU Astronomy Research Center, Astronomy Program, Department of Physics and Astronomy, Seoul National University, Seoul 08826, Republic of Korea}

\author{Clotilde Laigle}
\affiliation{Institut d’Astrophysique de Paris, UMR 7095, CNRS, Sorbonne Universit\'e, 98 bis boulevard Arago, F-75014 Paris, France}

\correspondingauthor{Myungshin Im}
\email{myungshin.im@gmail.com}

\begin{abstract}
It is yet to be understood how large-scale environments influence star formation activity in galaxy clusters. One recently proposed mechanism is that galaxy clusters can remain star-forming when fed by infalling groups and star-forming galaxies from large-scale structures surrounding them (the \textit{``web-feeding model"}). Using the COSMOS2020 catalog that has half a million galaxies with high accuracy ($\sigma_{\Delta z /1+z} \sim 0.01$) photometric redshifts, we study the relationship between star formation activities in galaxy clusters and their surrounding environment to test the web-feeding model. We first identify $68$ cluster candidates at $0.3 \leq z \leq 1.4$ with halo masses at $10^{13.0} - 10^{14.5}$ \SI{}{M_{\odot}}, and the surrounding large-scale structures (LSSs) with the friends-of-friends algorithm. We find that clusters with low fractions of quiescent galaxies tend to be connected with extended LSSs as expected in the web-feeding model. We also investigated the time evolution of the web-feeding trend using the IllustrisTNG cosmological simulation. Even though no clear correlation between the quiescent galaxy fraction of galaxy clusters and the significance of LSSs around them is found in the simulation, we verify that the quiescent galaxy fractions of infallers such as groups ($M_{200} \geq 10^{12}$ \SI{}{M_{\odot}}) and galaxies ($M_{200} < 10^{12}$ \SI{}{M_{\odot}}) is smaller than the quiescent fraction of cluster members and that infallers can lower the quiescent fraction of clusters. These results imply that cluster-to-cluster variations of quiescent galaxy fraction at $z \leq 1$ can at least partially be explained by feeding materials through cosmic webs to clusters.

\end{abstract}


\section{Introduction}\label{introduction}
\defcitealias{lee2019more}{L19}

As the largest gravitationally-bound objects in the universe, galaxy clusters are useful tools for constraining cosmological models of the universe. Galaxy clusters originate from the collapse of the overdensities in the initial density fluctuation field. These overdensities subsequently grew by accreting material from the large-scale structure. Thanks to their prominent scale in mass and size, galaxy clusters can offer unique laboratories to probe both the dynamical evolution of galaxies and gravitational models. Despite considerable progress in understanding galaxy clusters and their surrounding large-scale structures, it is not fully understood what factors play an important role in influencing star formation activity in galaxy clusters.

At low redshifts, galaxy clusters are known to have a higher fraction of red, early-type, and quiescent galaxies than in the field (e.g. \citealt{butcher1978evolution, 1993MNRAS.262..764A, 2007MNRAS.374..809D, 2007PhDT........39S}). While the fraction of quiescent galaxies declines in both cluster and field with redshift, the trend of high density regions having higher quiescent galaxy fraction than in field continues to $z \sim 1$ (e.g., \citealt{2001AJ....122.1861S, 2004ApJ...600..681B, 2004ApJ...601L..29H, 2004MNRAS.351.1151B, 2005MNRAS.362..268T,2007ApJS..173..315S}). Moreover, at higher redshifts, galaxy clusters have a wide range of quiescent galaxy fractions, which requires further explanation (e.g., \citealt{2012ApJ...746..188M, lee2015evolution, 2016ApJ...825..113D, 2017ApJ...847..134K}). 

To produce quiescent galaxies, a galaxy quenching mechanism is necessary. There are several quenching mechanisms that turn star-forming galaxies (in the blue cloud) into passive galaxies (in the red sequence) \citep{2004ApJ...600..681B,2004ApJ...608..752B}. Even though a clearer view of how quenching in galaxies takes place is still required, detailed explanations have been extensively built upon observational evidence (e.g., \citealt{peng2010mass,2022Univ....8..554A}). Mass quenching, also known as internal feedback, refers to all the internal processes linked to the galaxy mass, such as gas outflows driven by stellar winds or supernovae explosions (e.g., \citealt{larson1974dynamical, dekel1986origin, dalla2008simulating}). Also, several studies suggest that the AGN feedback from the central supermassive black hole (e.g., \citealt{croton2006many, fabian2012observational, fang2013link, cicone2014massive, bremer2018galaxy}) plays an important role in the mass quenching. On the other hand, environmental quenching is the physical process that stops star formation in these galaxies that are interacting with the surrounding area at a larger scale than the host halo. Environmental quenching includes hydrodynamical processes such as ram pressure stripping \citep{gunn1972infall}, and starvation or strangulation \citep{larson1980evolution}. Gravitational interactions through mergers, tidal interactions, and harassment can also trigger drastic changes in star formation (e.g. \citealt{moore1996galaxy, 2010MNRAS.405.1723S, 2015A&A...576A.103B}). In the local universe, the environment and mass effects on quenching can be separable to some extent, thanks to the richness of spectroscopic and photometric information used to constrain both galaxy redshift, stellar mass and stellar-to-halo mass ratio (e.g., \citealt{baldry2006galaxy, peng2010mass, kovavc2014zcosmos, balogh2016evidence, van2018stellar}), however, it is difficult to differentiate the two effects at higher redshift where data quality is poorer.

Because galaxies accrete the fuel for star formation from the cosmic web, it has naturally emerged as a potential factor to control quiescent galaxy fraction within galaxy clusters in the cosmological context. One proposed mechanism is \textit{Cosmic Web Detachment} (CWD), suggested by \citet{calvo2019galaxy}. According to the CWD, once the primordial filaments are detached or ruptured from the node, star formation starts to decline. This model aims at explaining how star formation is regulated across all mass ranges in a cosmological framework. The role of the cosmic web can be extended from galactic scales to larger scales. The filamentary structures replenish the galaxy cluster with star-forming galaxies, groups, and cold gas as they fall into the galaxy clusters through filaments. Previous studies have provided observational evidence that supports the enhanced star formation around the host cluster and nearby environment such as filaments (e.g., \citealt{bai2007ir, porter2007star, fadda2007starburst, koyama2008mapping, bai2009infrared, lubin2009observations, tanaka2009star,chung2010star, geach2011evolution, lemaux2012assembly, mahajan2012plunging, darvish2014cosmic, hung2016large, kleiner2017evidence, pintos2019evolution, einasto2020multiscale}). However, there are theoretical (e.g., \citealt{2018MNRAS.476.4877M, kraljic2020impact}) and observational (e.g., \citealt{2016MNRAS.457.2287A, laigle2018cosmos2015, 2018ApJ...852..142C, 2019MNRAS.483.3227K, 2021MNRAS.505.4920W}) works at odds with the trend. For example, \citet{2021MNRAS.501.4635S} points out that the quenching of galaxies specifically occurs at the edge of filaments. The coherent flow from vorticity-rich filaments (e.g., \citealt{2015MNRAS.446.2744L, 2022MNRAS.509.2707L}) impedes the gas transfer to the inner halo and lowers the efficiency of star formation. Therefore, the exact role of the cosmic web in regulating galaxy star formation still remains elusive. One important aspect is to conduct a meticulous comparison of various studies within a consistent mass and redshift range as the trend can vary significantly depending on the measured parameters and the scale under consideration.

Recently, \citet[hereafter \citetalias{lee2019more}]{lee2019more}, suggested the \textit{web-feeding model} that elucidates the variety levels of SF activity within clusters. By analyzing galaxies at $z \sim 1$ in the Ultra Deep Survey (UDS) field \citep{2007ASPC..379..163A}, \citetalias{lee2019more} found that member galaxies embedded within more extended structures tend to have a lower fraction of quiescent galaxies in comparison to those in isolated environments at a similar redshift. The correlation between quenched fraction and the size of connected large-scale led \citetalias{lee2019more} to propose that the enhanced star-forming activities in some of the overdensities at $z \sim 1$ are due to the inflow of gas and star-forming galaxies to the overdense areas from the surrounding large-scale environements\footnote{Here we confine the large-scale structures to several \SI{}{\mega pcs} probing the inter-cluster cosmic web.}. 

The main caveat of the \citetalias{lee2019more} is that the identified structures such as galaxy clusters and surrounding filaments are susceptible to line-of-sight contamination due to large photometric redshift uncertainties of about $0.028(1+z)$. This could lead to erroneous associations of galaxy clusters with the large-scale structure and systematic errors in the quiescent fraction of galaxies due to the misidentification of cluster members. Additionally, 
the result could be susceptible to cosmic variance \citep{2011ApJ...731..113M}. Therefore, examining the web-feeding model using an independent field with improved photometric redshift accuracy is highly desired.

In this paper, we will test the web-feeding model with the COSMOS2020 data \citep{weaver2022cosmos2020}. As described in the next section, the newly released COSMOS2020 data provides photometric redshifts that are several times more accurate than those used in \citetalias{lee2019more} and also contains tens of thousands of spectroscopic redshifts. Furthermore, the COSMOS field \citep{scoville2007cosmic} is nearly twice larger than the field of view of the UDS field. Thus, with the COSMOS2020 data, it is possible to significantly improve the analysis of \citetalias{lee2019more}. Moreover, we will also investigate the time evolution of large-scale cosmic web feeding and the respective effects of infalling structures using IllustrisTNG 300-1 simulation \citep{springel2018first, nelson2018first}. Throughout this work, we adopt the standard $\Lambda$CDM cosmology ($\Omega_{m}$, $\Omega_{\lambda}$) = ($0.3, 0.7$) and $H_{0} = 70$ \SI{}{km \second^{-1} \mega pc^{-1}} and AB magnitude system \citep{1974ApJS...27...21O}.

\section{Data}

The Cosmic Evolution Survey (COSMOS) is a deep multi-wavelength survey of a \SI{2}{deg^{2}} of the sky centered at RA of 10:00:28.8 and Dec of +02:12:21.0 \citep{scoville2007cosmic}. It boasts data from the X-ray to the radio, including the Hubble Space Telescope and the Chandra X-ray images for studying distant galaxies at high spatial resolution. COSMOS also includes a multitude of ground-based imaging and spectroscopic data. In particular, it contains narrow and medium band images covering the optical to near-infrared including NB711, NB816, and 12 medium-bands from Subaru Suprime-Cam \citep{taniguchi2007cosmic, taniguchi2015subaru} and NB118 from UltraVISTA survey \citep{mccracken2015probing, moneti2019fourth}. Moreover, ultra-deep images such as $J^{\textrm{UD}}$, $H^{\textrm{UD}}$, and $K^{\textrm{UD}}$ reach $3\sigma$ depths in \ang{;;3} diameter apertures of 25.9, 25.5, and 25.2 mags respectively, which are useful for accurately determining photometric redshifts (see \citealt{weaver2022cosmos2020} for more details). More importantly, about $20,000$ targets of spectroscopic redshifts have been obtained in this field largely from the zCOSMOS survey \citep{lilly2007zcosmos} and VIMOS Ultra Deep Survey (VUDS; \citealt{le2015vimos}), making it possible to test photometric redshifts thoroughly.

In this study, we use the most up-to-date publicly released catalog produced by the team, namely COSMOS2020 \citep{weaver2022cosmos2020}. Since the last public catalog in 2015 \citep{laigle2016cosmos2015}, new photometric and spectroscopic data has been added including ultra-deep optical data from Hyper Suprime-Cam (HSC) Subaru Strategic Program (SSP) PDR2 \citep{aihara2019second}, Visible Infrared Survey Telescope for Astronomy (VISTA) DR4, and Spitzer IRAC data \citep{2018ApJS..237...39A}. With these additions, the number of detected sources doubled, and homogeneity in photometry and astrometry was improved significantly. As a result, \citet{weaver2022cosmos2020} suggests that COSMOS2020 contains the most reliable photometric redshifts of galaxies in the COSMOS field at present. The photometric redshift accuracy is only sub-percent for bright sources ($i < 21$) and $5\%$ at $25 < i < 27$. 

There are two versions of the COSMOS2020 catalog provided: \texttt{CLASSIC} and \texttt{FARMER}. The source detection in the \texttt{CLASSIC} catalog is performed using \texttt{SExtractor} \citep{bertin1996sextractor}. On the other hand, the \textsc{FARMER} catalog utilizes \texttt{The Tractor} \citep{lang2016tractor} that has been developed to perform profile-fitting photometry. This model-based code enables photometry of the detected sources free from blending with close objects and from PSF-homogenization while suffering from different resolution regimes and failure of convergence for either extremely bright or dense sources. The catalogs obtained from two different photometric extraction codes are in good agreement overall, but the choice of the catalog should depend on the study's specific objectives.

For photometric redshift and SED fitting, the results from two separate codes are also available; \texttt{LePhare} \citep{2002MNRAS.329..355A, ilbert2006accurate} and \texttt{EAZY} \citep{brammer2008eazy}. Compared with spectroscopic redshifts in the COSMOS field, the Normalized Median Absolute Deviation (NMAD, \citealt{hoaglin1983understanding}) of photometric redshift is of the order of $0.01(1+z)$ at $i < 22.5$ and better than $0.25(1+z)$ at fainter magnitudes for both cases. Even though the precision of photometric redshifts is similar between both cases, \texttt{FARMER} has its advantages at fainter magnitudes (lower NMAD), while \texttt{CLASSIC} presents better validity at brighter sources (see Figure 13 and 15 in \citealt{weaver2022cosmos2020}). Given our primary precondition for this study is to identify reliable galaxy cluster candidates and surrounding large-scale structures at relatively high redshift, we adopt a combination of \texttt{FARMER} and \texttt{LePhare}, which shows the smaller fraction of catastrophic failure $\eta$, the ratio of deviant galaxies from their spectroscopic redshift by $\Delta z > 0.15(1+z_{spec})$ with similar precision.

\subsection{Photometric Redshift Uncertainties}

Because our goal is to find reliable cluster members and minimize the contamination from line-of-sight interlopers, we need to confine the photometric redshift uncertainty to an appropriate level. The threshold for photometric redshift uncertainty should not be too strict to avoid excluding the high-redshift region but also not too loose to avoid contaminating the cluster members in the foreground or background direction. Previous studies \citep{cooper2005measuring, malavasi2016reconstructing, darvish2017cosmic} have verified that photometric redshifts with uncertainties of $\sigma_{\Delta z /1+z} \sim 0.01$ can reliably build the density field. In the following analysis, we adopt the $0.01(1+z)$ as a fiducial value to determine limiting quantities such as maximum redshift and limiting stellar mass.

\begin{figure}
    \centering
    \includegraphics[width = 0.45\textwidth]{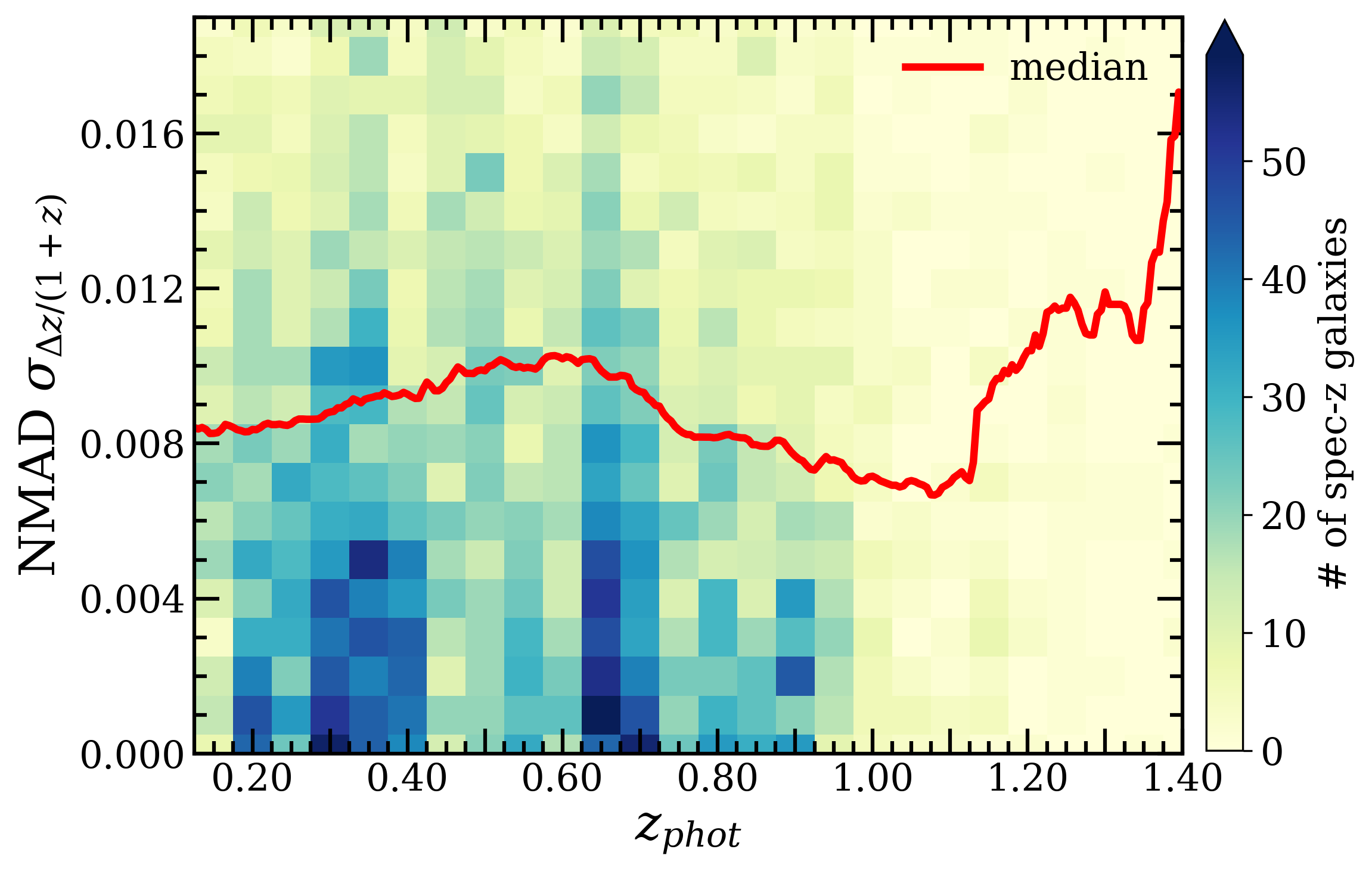}
    \caption{Photometric redshift uncertainty (NMAD, $\sigma_{\Delta z /1+z}$) as a function of photometric redshift (red line). The uncertainty is calculated by comparing photometric redshift derived from \texttt{LePhare} with publicly available spectroscopic redshift catalog \citep{lilly2007zcosmos}. The background 2D  histogram shows the population of galaxies within the photometric redshift and uncertainty bins.}
    \label{fig:photz_uncertainty}
\end{figure}

To examine the COSMOS2020 photometric redshift accuracy, we compared the photometric redshifts from the COSMOS2020 and matched spectroscopic redshifts of $8,562$ galaxies from zCOSMOS \citep{lilly2007zcosmos}. As shown in Figure \ref{fig:photz_uncertainty}, we find that the NMAD $\sigma_{\Delta z /1+z}$ is of the order of $0.01$ up to a redshift of $z < 1.4$. This result is consistent with the result in \citet{weaver2022cosmos2020} where they found the same order of NMAD at $17 < i < 24$  (See Figure 17 of their paper for more details). This NMAD $\sigma_{\Delta z /1+z}$ corresponds to the galaxies matched with the COSMOS2020 data, mostly brighter than the limiting magnitudes from various surveys of COSMOS2020. Note that this photometric redshift is applicable to the brighter galaxies with spectroscopic redshifts available. For example, the spectroscopic redshift sample has a mean and standard deviation in HSC $i$ band magnitudes of 21.4 and 0.9 mags, respectively, while the photometric redshift sample, used throughout the paper, has $23.7$ and $1.5$ mags. Although our sample includes faint galaxies, the majority ($\gtrsim80\%$) of the sample is brighter than $24.9$ mag at which the NMAD values are of the order of $0.02$ to $0.03$ \citep{weaver2022cosmos2020}. Such an order of uncertainties, $3\times0.01(1+z)$ is taken into account when finding clusters in Section \ref{sec:galaxy_cluster_selection}.

\subsection{Mass Complete Sample}\label{section:mass_complete_sample}

To avoid the bias arising by missing faint low-mass galaxies, we construct the mass complete sample by following the empirical procedure adopted by \citet{pozzetti2010zcosmos} and \citet{ilbert2013mass}. The idea of this approach is to transform the detection limit of a survey, represented as the apparent magnitude $m_{\textrm{lim}}$, into the observable stellar mass limit $M_{\ast, \textrm{lim}}$ as a function of redshift. We use  $m_{\textrm{lim}}$ of the IRAC channel 1 from the CANDELS-COSMOS catalog \citep{nayyeri2017candels}. The $m_{\textrm{lim}}$ in the IRAC channel 1 is set to $26$ mag, corresponding to the $3\sigma$ depth of $26.4/25.7$ mags in the aperture of \ang{;;2}$/$ \ang{;;3} \citep{davidzon2017cosmos2015, weaver2022cosmos2020}. Then, we convert the apparent magnitude $m_{i}$ of the $i$-th galaxy into the stellar mass, which is scaled by an empirical mass-to-light ratio $10^{-0.4(m_{i} - m_{\textrm{lim}})}$. In the next step, we determine the 95th percentile stellar mass completeness limit, which is defined as the 95th percentile of the smallest mass at the central redshift of each redshift bin with a step size of $\Delta z = 0.05$.

\begin{figure}
    \centering
    \includegraphics[width = 0.45\textwidth]{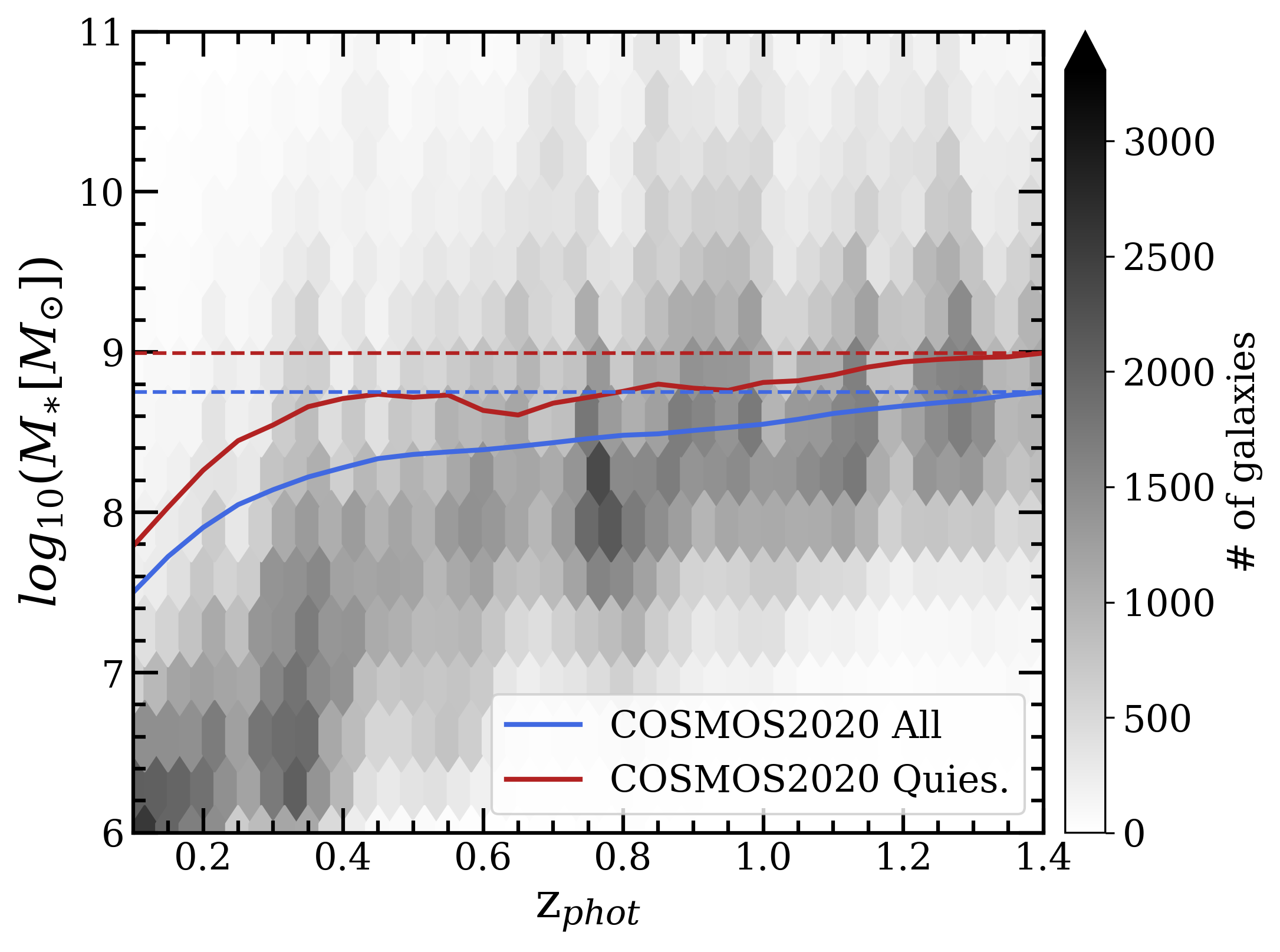}
    \caption{Mass complete limit as a function of photometric redshift. The blue solid line represents the mass complete limit of all types of galaxies while the red solid line is for quiescent galaxies only. The blue and red dashed lines indicate the mass completeness limits for all types and quiescent galaxies at = 1.4. We used galaxies exceeding the stellar mass limit as indicated by the blue dashed line. The background 2D histogram stands for the number of galaxies in a given redshift and stellar mass bin.}
    \label{fig:mass_complete}
\end{figure}

The stellar mass limit at z = 1.4, where the uncertainty of photometric redshift is as low as $0.01(1+z)$ to reliably build density structures, is $10^{8.75}$ \SI{}{M_{\odot}} for all types of galaxies and $10^{8.99}$ \SI{}{M_{\odot}} for quiescent galaxies selected based on Eq. {\ref{eq:quiescent_flag}} as shown in Figure \ref{fig:mass_complete}. When we construct the density field and find galaxy clusters, we apply this stellar mass cut. However, it is possible that this selection is not complete for low-mass quiescent galaxies. Therefore, we adopt the mass limit of $10^{8.99}$ \SI{}{M_{\odot}} when we calculate the quiescent galaxy fraction (see Section \ref{subsec:galaxy_evolution_from_sf_to_q}). Compared with \citealt{2023A&A...677A.184W}, we confirmed that our mass completeness limit is nearly consistent with the 70$\%$ mass completeness limit of CANDELS-COSMOS sources ($10^{8.57}$ \SI{}{M_{\odot}} for all types of galaxies and $10^{8.91}$ \SI{}{M_{\odot}} for quiescent galaxies).

With the information obtained from the aforementioned calculation, we select sources that are flagged as galaxies (\texttt{lp\_type = 0}), outside the bright source mask (\texttt{FLAG\_COMBINED = 0}), and more massive than the mass complete limit of $10^{8.75}$ \SI{}{M_{\odot}}. By imposing the flag condition \texttt{FLAG\_COMBINED = 0} obtained from combining the bright source masks in the UltraVista \citep{2012A&A...544A.156M}, HSC-SSP \citep{2018PASJ...70S...7C}, and SuprimeCam \citep{taniguchi2007cosmic, taniguchi2015subaru} regions, we can avoid the data with unreliable photometry or partial coverage. Also, we limit our study to $z \leq 1.4$ to construct the reliable density field using accurate $\sigma_{\Delta z /1+z} \sim 0.01$. The total number of galaxies after we applied the source flags, the stellar mass cut, and the photometric redshift cut is $110,409$.

\subsection{Galaxy Cluster Selection}\label{sec:galaxy_cluster_selection}

Galaxy clusters are identified as overdense regions in the density field \citep{kang2015massive, lee2015evolution}. To construct the density field, we divide the galaxy sample into multiple redshift bins from $z = 0.1$ to $1.4$ with a step size of $\Delta z = 0.01$. The number of galaxies in each bin nearly uniformly increases from $\sim$100 at $z=0.1$ to $\sim$8000 at $z=1.4$. Here, the step size is determined as the value comparable to the photometric redshift accuracy $\sigma_{\Delta z /1+z}$. For galaxy redshifts, we use photometric redshifts except when spectroscopic redshifts are available from zCOSMOS \citep{lilly2007zcosmos}. Then, in each redshift bin, we count the number of galaxies within a search radius of \SI{700}{\kilo pc} at every point that is spaced at \SI{100}{\kilo pc}. A convolution radial scale of 700 kpc is chosen to probe structures slightly smaller than a typical galaxy cluster. This value lies within the range of typical filter sizes of density maps $0.5-1.0$ Mpc used to find cluster candidates (e.g., \citealt{2000AJ....119...12G, 2009ApJ...691L..33K, 2018A&A...613A..67S}). We select galaxy cluster candidates with a surface number density exceeding $4$ times the standard deviation from the average number density at a given redshift. Adopting the 4$\sigma$ threshold, as done by \citetalias{lee2019more}, allows us to compare our results consistently and to identify intermediate-mass overdensities where the web-feeding trend is expected to appear. Our selection of the galaxy cluster candidates is based on the following conditions. (1) Connected $4\sigma$ level overdense grid points should be more than 10 points; (2) Overdensities should be linked along the line-of-sight over at least three redshift bins. The condition of the number of connected points is imposed to sample overdense regions to the approximate size of galaxy clusters corresponding to $R_{200}$ $\sim$ \SI{1}{\mega pc}. Furthermore, the choice of more than three redshift grids linked along the ling-of-sight aims to detect as many candidates as possible and to avoid including the falsely overlapping structures in photometric redshifts in our sample. The completeness of this method is further investigated at the end of this section.

To determine member galaxies, the initial center coordinate (RA, Dec, z) of a cluster candidate is estimated as the number density-weighted average of the coordinates for all the connected grid points. Along the line-of-sight direction, we apply a conservative condition to protect member galaxies from contamination derived from the photometric redshift uncertainties and only select galaxies within a given redshift bin $|z| \leq z_{\textrm{grid}} \pm \sigma_{\Delta z /1+z}(1+z)$, where $z_{\textrm{grid}}$ is a redshift of a given redshift bin. Then, we calculate the transversal distance distribution of the galaxies from the initial center. The 2D distance distribution from the initial center shows a bell-like shape and we therefore fit the distribution with a Gaussian distribution. The transversal cluster boundary from the center is then determined as $3\sigma$ of the Gaussian distribution and members of a cluster candidate are defined as galaxies within the corresponding radius. As for final member candidates, we exclude galaxies whose spectroscopic redshifts (1) differ from photometric redshifts more than \SI{15}{\%} or (2) are outside the $\pm 3 \sigma_{\Delta z /1+z}(1+z)$ range from the redshift of the cluster center. Finally, we re-calculate the cluster's central position and redshift by the mass-weighted mean of member galaxies. As a result, $109$ cluster candidates are identified. Furthermore, we exclude the candidates that are near the bright source masks and(or) survey edges $(39/109)$ or that are largely elongated along the line-of-sight direction $(2/109)$. The remaining number of candidates becomes $68$. These clusters and their properties are listed in Table \ref{tab:clusters}.

\startlongtable
\begin{deluxetable*}{ccccccccc}
\tabletypesize{\scriptsize}
\tablewidth{0pt}
\tablecaption{Galaxy cluster candidates found in the COSMOS field. The full table is available online.\label{tab:clusters}}

\tablehead{
\colhead{R.A. (J2000)} & \colhead{dec. (J2000)} & \colhead{$z_{\textrm{phot}}$} & \colhead{$\log{(M_{200}/M_{\odot})}$}&\colhead{$N_{\textrm{mem}}$} & \colhead{$N_{\textrm{outlier}} / N_{\textrm{spec}}$} & \colhead{\textit{FoF fraction}} & \colhead{\textit{QF}} & \colhead{ID}
}

\startdata
150.045 & 2.216 & ${0.266}^{+0.008}_{-0.007}$& ${13.64}^{+0.12}_{-0.06}$& 23 & 2/12 & ${0.045}^{+0.009}_{-0.004}$& ${0.44}^{+0.19}_{-0.17}$& - \\
150.306 & 2.016 & ${0.309}^{+0.008}_{-0.007}$& ${13.12}^{+0.15}_{-0.17}$& 23 & 1/9 & ${0.037}^{+0.013}_{-0.009}$& ${0.40}^{+0.09}_{-0.10}$& 20077 \\
150.189 & 1.759 & ${0.333}^{+0.009}_{-0.005}$& ${13.47}^{+0.11}_{-0.01}$& 53 & 4/23 & ${0.016}^{+0.000}_{-0.003}$& ${0.68}^{+0.23}_{-0.02}$& 20029 \\
149.945 & 2.601 & ${0.333}^{+0.010}_{-0.006}$& ${13.39}^{+0.52}_{-0.27}$& 29 & 2/15 & ${0.024}^{+0.001}_{-0.009}$& ${0.50}^{+0.03}_{-0.05}$& 30311 \\
150.485 & 2.056 & ${0.431}^{+0.007}_{-0.006}$& ${13.50}^{+0.15}_{-0.12}$& 31 & 6/16 & ${0.053}^{+0.017}_{-0.000}$& ${0.56}^{+0.03}_{-0.06}$& 30315 \\
149.964 & 2.207 & ${0.435}^{+0.005}_{-0.007}$& ${13.55}^{+0.07}_{-0.09}$& 23 & 5/18 & ${0.007}^{+0.000}_{-0.000}$& ${0.91}^{+0.02}_{-0.08}$& 20088 \\
150.112 & 2.562 & ${0.505}^{+0.008}_{-0.008}$& ${13.20}^{+0.07}_{-0.04}$& 70 & 4/25 & ${0.024}^{+0.000}_{-0.021}$& ${0.43}^{+0.03}_{-0.05}$& 20137 \\
150.223 & 1.815 & ${0.543}^{+0.003}_{-0.009}$& ${13.65}^{+0.15}_{-0.00}$& 83 & 1/35 & ${0.064}^{+0.001}_{-0.010}$& ${0.35}^{+0.04}_{-0.04}$& 20289 \\
150.133 & 1.860 & ${0.547}^{+0.009}_{-0.012}$& ${13.89}^{+0.21}_{-0.02}$& 60 & 4/27 & ${0.066}^{+0.013}_{-0.008}$& ${0.33}^{+0.13}_{-0.07}$& - \\
149.915 & 2.523 & ${0.602}^{+0.008}_{-0.005}$& ${13.31}^{+0.07}_{-0.06}$& 80 & 3/4 & ${0.025}^{+0.005}_{-0.000}$& ${0.57}^{+0.06}_{-0.11}$& - \\
149.729 & 1.836 & ${0.597}^{+0.010}_{-0.010}$& ${13.37}^{+0.20}_{-0.05}$& 26 & 3/6 & ${0.089}^{+0.032}_{-0.019}$& ${0.28}^{+0.05}_{-0.10}$& - \\
150.503 & 2.454 & ${0.626}^{+0.009}_{-0.008}$& ${13.44}^{+0.26}_{-0.06}$& 47 & 4/11 & ${0.080}^{+0.003}_{-0.019}$& ${0.32}^{+0.18}_{-0.05}$& - \\
149.602 & 1.892 & ${0.655}^{+0.005}_{-0.004}$& ${13.57}^{+0.20}_{-0.06}$& 72 & 0/12 & ${0.031}^{+0.018}_{-0.000}$& ${0.25}^{+0.13}_{-0.08}$& - \\
150.151 & 2.499 & ${0.658}^{+0.005}_{-0.006}$& ${13.39}^{+0.24}_{-0.04}$& 47 & 1/10 & ${0.033}^{+0.001}_{-0.014}$& ${0.29}^{+0.04}_{-0.10}$& 20035 \\
149.927 & 2.104 & ${0.663}^{+0.005}_{-0.005}$& ${13.59}^{+0.19}_{-0.07}$& 57 & 3/18 & ${0.055}^{+0.003}_{-0.000}$& ${0.29}^{+0.04}_{-0.08}$& - \\
150.058 & 2.611 & ${0.675}^{+0.007}_{-0.015}$& ${13.70}^{+0.11}_{-0.16}$& 59 & 2/15 & ${0.056}^{+0.000}_{-0.030}$& ${0.33}^{+0.07}_{-0.03}$& 10215 \\
150.086 & 2.192 & ${0.697}^{+0.008}_{-0.005}$& ${13.78}^{+0.22}_{-0.08}$& 31 & 1/9 & ${0.010}^{+0.036}_{-0.006}$& ${0.43}^{+0.08}_{-0.13}$& 10216 \\
150.052 & 2.308 & ${0.717}^{+0.009}_{-0.010}$& ${13.63}^{+0.06}_{-0.03}$& 38 & 0/13 & ${0.107}^{+0.018}_{-0.001}$& ${0.22}^{+0.06}_{-0.10}$& - \\
150.039 & 2.649 & ${0.792}^{+0.008}_{-0.005}$& ${13.22}^{+0.02}_{-0.10}$& 28 & 1/4 & ${0.013}^{+0.004}_{-0.000}$& ${0.40}^{+0.08}_{-0.07}$& - \\
150.532 & 2.160 & ${0.834}^{+0.008}_{-0.007}$& ${13.78}^{+0.42}_{-0.01}$& 170 & 1/30 & ${0.080}^{+0.001}_{-0.022}$& ${0.27}^{+0.03}_{-0.02}$& - \\
150.688 & 2.418 & ${0.825}^{+0.007}_{-0.006}$& ${13.60}^{+0.21}_{-0.12}$& 33 & 0/1 & ${0.027}^{+0.000}_{-0.014}$& ${0.30}^{+0.02}_{-0.02}$& - \\
149.651 & 2.386 & ${0.841}^{+0.005}_{-0.008}$& ${13.95}^{+0.08}_{-0.04}$& 111 & 2/8 & ${0.044}^{+0.011}_{-0.000}$& ${0.44}^{+0.03}_{-0.02}$& 30231 \\
150.374 & 2.141 & ${0.840}^{+0.006}_{-0.007}$& ${13.76}^{+0.05}_{-0.02}$& 84 & 2/17 & ${0.097}^{+0.024}_{-0.000}$& ${0.26}^{+0.04}_{-0.05}$& - \\
149.553 & 2.421 & ${0.837}^{+0.005}_{-0.007}$& ${13.78}^{+0.07}_{-0.02}$& 31 & 0/5 & ${0.053}^{+0.004}_{-0.009}$& ${0.44}^{+0.03}_{-0.03}$& 20106 \\
150.453 & 2.142 & ${0.861}^{+0.009}_{-0.015}$& ${13.92}^{+0.14}_{-0.06}$& 64 & 3/10 & ${0.065}^{+0.007}_{-0.020}$& ${0.25}^{+0.06}_{-0.03}$& - \\
150.553 & 2.197 & ${0.847}^{+0.008}_{-0.007}$& ${13.69}^{+0.07}_{-0.02}$& 54 & 0/9 & ${0.079}^{+0.006}_{-0.012}$& ${0.35}^{+0.03}_{-0.06}$& - \\
149.985 & 2.321 & ${0.860}^{+0.008}_{-0.008}$& ${14.12}^{+0.10}_{-0.02}$& 40 & 6/10 & ${0.039}^{+0.027}_{-0.007}$& ${0.58}^{+0.03}_{-0.03}$& - \\
150.220 & 2.287 & ${0.870}^{+0.009}_{-0.007}$& ${13.86}^{+0.12}_{-0.16}$& 42 & 1/7 & ${0.047}^{+0.008}_{-0.000}$& ${0.39}^{+0.04}_{-0.09}$& 20135 \\
149.934 & 2.406 & ${0.886}^{+0.004}_{-0.011}$& ${13.77}^{+0.34}_{-0.01}$& 105 & 2/20 & ${0.023}^{+0.019}_{-0.000}$& ${0.48}^{+0.11}_{-0.04}$& 20187 \\
150.088 & 2.533 & ${0.888}^{+0.006}_{-0.008}$& ${13.89}^{+0.24}_{-0.02}$& 111 & 1/17 & ${0.062}^{+0.011}_{-0.006}$& ${0.40}^{+0.11}_{-0.06}$& 10208 \\
149.552 & 2.003 & ${0.884}^{+0.005}_{-0.006}$& ${13.64}^{+0.31}_{-0.31}$& 45 & 0/3 & ${0.009}^{+0.007}_{-0.002}$& ${0.29}^{+0.09}_{-0.03}$& 20143 \\
149.925 & 2.642 & ${0.889}^{+0.002}_{-0.007}$& ${13.99}^{+0.06}_{-0.01}$& 196 & 3/30 & ${0.075}^{+0.000}_{-0.007}$& ${0.21}^{+0.02}_{-0.02}$& - \\
149.671 & 2.257 & ${0.911}^{+0.003}_{-0.005}$& ${13.48}^{+0.04}_{-0.01}$& 47 & 0/6 & ${0.009}^{+0.001}_{-0.000}$& ${0.17}^{+0.02}_{-0.03}$& - \\
149.976 & 2.341 & ${0.933}^{+0.003}_{-0.005}$& ${14.07}^{+0.07}_{-0.06}$& 205 & 4/48 & ${0.098}^{+0.000}_{-0.000}$& ${0.41}^{+0.03}_{-0.04}$& 30172 \\
150.261 & 2.075 & ${0.930}^{+0.007}_{-0.007}$& ${13.62}^{+0.15}_{-0.01}$& 72 & 2/8 & ${0.076}^{+0.002}_{-0.000}$& ${0.21}^{+0.01}_{-0.01}$& - \\
150.159 & 2.192 & ${0.928}^{+0.003}_{-0.003}$& ${13.68}^{+0.12}_{-0.00}$& 45 & 1/10 & ${0.117}^{+0.000}_{-0.000}$& ${0.18}^{+0.02}_{-0.01}$& - \\
150.085 & 2.193 & ${0.932}^{+0.005}_{-0.006}$& ${13.78}^{+0.04}_{-0.03}$& 51 & 1/6 & ${0.114}^{+0.002}_{-0.001}$& ${0.26}^{+0.04}_{-0.02}$& - \\
150.030 & 2.201 & ${0.940}^{+0.004}_{-0.004}$& ${13.92}^{+0.13}_{-0.11}$& 128 & 4/17 & ${0.117}^{+0.007}_{-0.000}$& ${0.22}^{+0.03}_{-0.05}$& 10281 \\
150.036 & 2.302 & ${0.930}^{+0.006}_{-0.004}$& ${13.71}^{+0.13}_{-0.03}$& 51 & 1/12 & ${0.110}^{+0.000}_{-0.003}$& ${0.16}^{+0.04}_{-0.00}$& - \\
149.652 & 2.343 & ${0.960}^{+0.009}_{-0.006}$& ${13.63}^{+0.02}_{-0.13}$& 109 & 2/8 & ${0.078}^{+0.010}_{-0.000}$& ${0.29}^{+0.05}_{-0.04}$& 30296 \\
149.646 & 2.222 & ${0.960}^{+0.006}_{-0.008}$& ${13.54}^{+0.01}_{-0.02}$& 69 & 0/4 & ${0.084}^{+0.008}_{-0.000}$& ${0.29}^{+0.05}_{-0.02}$& 20161 \\
149.494 & 2.012 & ${0.988}^{+0.006}_{-0.006}$& ${14.00}^{+0.13}_{-0.02}$& 106 & 0/3 & ${0.103}^{+0.005}_{-0.000}$& ${0.16}^{+0.01}_{-0.05}$& - \\
149.748 & 2.267 & ${1.017}^{+0.002}_{-0.004}$& ${13.99}^{+0.14}_{-0.01}$& 211 & 14/35 & ${0.121}^{+0.009}_{-0.000}$& ${0.28}^{+0.01}_{-0.02}$& - \\
149.972 & 1.672 & ${1.028}^{+0.006}_{-0.003}$& ${13.40}^{+0.11}_{-0.06}$& 40 & 1/3 & ${0.024}^{+0.001}_{-0.000}$& ${0.13}^{+0.01}_{-0.02}$& - \\
150.704 & 2.312 & ${1.080}^{+0.010}_{-0.009}$& ${13.57}^{+0.08}_{-0.08}$& 111 & 0/4 & ${0.098}^{+0.016}_{-0.007}$& ${0.18}^{+0.04}_{-0.03}$& 20150 \\
150.636 & 2.410 & ${1.102}^{+0.004}_{-0.007}$& ${13.31}^{+0.11}_{-0.01}$& 28 & 1/1 & ${0.066}^{+0.000}_{-0.016}$& ${0.17}^{+0.01}_{-0.02}$& - \\
150.541 & 2.550 & ${1.136}^{+0.007}_{-0.005}$& ${13.91}^{+0.03}_{-0.20}$& 89 & 3/4 & ${0.059}^{+0.001}_{-0.000}$& ${0.26}^{+0.04}_{-0.01}$& - \\
150.437 & 2.542 & ${1.128}^{+0.007}_{-0.007}$& ${13.43}^{+0.06}_{-0.08}$& 38 & 2/3 & ${0.053}^{+0.000}_{-0.002}$& ${0.17}^{+0.02}_{-0.03}$& - \\
150.351 & 1.953 & ${1.148}^{+0.005}_{-0.008}$& ${13.55}^{+0.09}_{-0.01}$& 117 & 8/11 & ${0.035}^{+0.000}_{-0.006}$& ${0.10}^{+0.03}_{-0.03}$& - \\
149.907 & 2.673 & ${1.141}^{+0.004}_{-0.005}$& ${13.37}^{+0.13}_{-0.04}$& 39 & 2/3 & ${0.012}^{+0.002}_{-0.000}$& ${0.10}^{+0.02}_{-0.05}$& - \\
150.199 & 1.899 & ${1.181}^{+0.007}_{-0.005}$& ${13.41}^{+0.18}_{-0.30}$& 35 & 2/3 & ${0.083}^{+0.000}_{-0.018}$& ${0.22}^{+0.03}_{-0.02}$& - \\
150.122 & 1.984 & ${1.187}^{+0.005}_{-0.004}$& ${13.58}^{+0.01}_{-0.03}$& 273 & 19/51 & ${0.092}^{+0.017}_{-0.000}$& ${0.25}^{+0.06}_{-0.02}$& - \\
150.098 & 2.032 & ${1.190}^{+0.002}_{-0.001}$& ${13.83}^{+0.04}_{-0.00}$& 117 & 12/20 & ${0.090}^{+0.000}_{-0.000}$& ${0.29}^{+0.01}_{-0.01}$& - \\
149.896 & 2.237 & ${1.187}^{+0.001}_{-0.005}$& ${13.67}^{+0.02}_{-0.06}$& 43 & 1/7 & ${0.034}^{+0.008}_{-0.000}$& ${0.13}^{+0.01}_{-0.02}$& - \\
149.998 & 2.664 & ${1.213}^{+0.011}_{-0.014}$& ${13.65}^{+0.12}_{-0.10}$& 27 & 0/3 & ${0.014}^{+0.007}_{-0.005}$& ${0.48}^{+0.15}_{-0.04}$& 20130 \\
149.700 & 2.014 & ${1.236}^{+0.007}_{-0.007}$& ${13.53}^{+0.06}_{-0.09}$& 204 & 5/19 & ${0.095}^{+0.000}_{-0.019}$& ${0.05}^{+0.02}_{-0.02}$& - \\
149.727 & 2.008 & ${1.233}^{+0.004}_{-0.007}$& ${13.59}^{+0.12}_{-0.07}$& 292 & 8/35 & ${0.111}^{+0.001}_{-0.019}$& ${0.08}^{+0.02}_{-0.02}$& - \\
150.586 & 1.963 & ${1.271}^{+0.005}_{-0.004}$& ${13.85}^{+0.05}_{-0.05}$& 106 & 2/4 & ${0.037}^{+0.000}_{-0.000}$& ${0.11}^{+0.02}_{-0.05}$& - \\
149.995 & 2.685 & ${1.290}^{+0.006}_{-0.010}$& ${13.62}^{+0.16}_{-0.13}$& 49 & 0/0 & ${0.129}^{+0.026}_{-0.009}$& ${0.10}^{+0.02}_{-0.01}$& - \\
150.247 & 2.698 & ${1.275}^{+0.003}_{-0.007}$& ${13.69}^{+0.17}_{-0.01}$& 94 & 0/4 & ${0.159}^{+0.000}_{-0.039}$& ${0.04}^{+0.02}_{-0.00}$& 20174 \\
149.950 & 2.547 & ${1.290}^{+0.004}_{-0.006}$& ${13.38}^{+0.07}_{-0.11}$& 46 & 2/4 & ${0.064}^{+0.015}_{-0.001}$& ${0.03}^{+0.00}_{-0.02}$& - \\
149.945 & 2.652 & ${1.294}^{+0.003}_{-0.008}$& ${13.43}^{+0.06}_{-0.04}$& 205 & 2/7 & ${0.083}^{+0.000}_{-0.017}$& ${0.05}^{+0.02}_{-0.00}$& - \\
149.947 & 2.634 & ${1.298}^{+0.003}_{-0.009}$& ${13.57}^{+0.04}_{-0.03}$& 89 & 1/4 & ${0.097}^{+0.000}_{-0.017}$& ${0.06}^{+0.03}_{-0.01}$& - \\
149.884 & 2.674 & ${1.364}^{+0.014}_{-0.006}$& ${13.28}^{+0.40}_{-0.29}$& 37 & 0/1 & ${0.056}^{+0.011}_{-0.011}$& ${0.04}^{+0.01}_{-0.07}$& - \\
149.817 & 2.017 & ${1.345}^{+0.007}_{-0.003}$& ${13.40}^{+0.03}_{-0.03}$& 23 & 6/7 & ${0.026}^{+0.000}_{-0.006}$& ${0.10}^{+0.00}_{-0.00}$& - \\
149.815 & 1.888 & ${1.395}^{+0.008}_{-0.005}$& ${13.56}^{+0.03}_{-0.09}$& 46 & 3/5 & ${0.038}^{+0.006}_{-0.000}$& ${0.08}^{+0.00}_{-0.05}$& 20134 \\
150.220 & 1.806 & ${1.393}^{+0.006}_{-0.005}$& ${13.41}^{+0.18}_{-0.04}$& 48 & 3/5 & ${0.037}^{+0.000}_{-0.010}$& ${0.05}^{+0.02}_{-0.01}$& - \\
149.856 & 2.125 & ${1.397}^{+0.003}_{-0.010}$& ${13.62}^{+0.01}_{-0.03}$& 25 & 4/6 & ${0.025}^{+0.006}_{-0.000}$& ${0.11}^{+0.02}_{-0.04}$& - \\
\enddata 

\tablecomments{We list (1-3) the center of R.A., dec., photometric redshift, (4) the halo mass estimated from the scaling relation with X-ray groups, (5) the number of member galaxies, (6) the number of photometric redshift outliers over the number of members with spectroscopic redshifts, (7) FoF fraction, (8) Quiescent galaxy fraction, and (9) the \texttt{ID\_COSMOS} in X-ray galaxy group catalog \citep{gozaliasl2019chandra} if the cluster candidates are matched within \SI{1}{h^{-1}\mega pc} and $|\Delta z| \leq 0.03(1+z)$.}
\end{deluxetable*}

To verify the reliability of the identified galaxy clusters, we use the lightcone mock catalog \citep{merson2013lightcone} from the Millennium simulation \citep{springel2005cosmological} and GALFORM semi-analytic model \citep{cole2000hierarchical, bower2006breaking}. To reproduce a field similar to the COSMOS, we define a $1.4 \times 1.4$ \SI{}{deg^{2}} area and use galaxies more massive than the stellar mass of $10^{8.75}$ \SI{}{M_{\odot}}. Moreover, we add photometric redshift errors following the Gaussian distribution with the standard deviation $\sigma$ that corresponds to the photometric redshift uncertainty at the observed redshifts. We apply the same cluster-finding method as the cluster search for the COSMOS2020 data but with a different number of the least connected redshift bins and compare the found cluster candidates to estimate the completeness of this method. There are $674$ ($14$) halos more massive than $10^{13}$ ($10^{14}$) \SI{}{M_{\odot}} at $z \leq 1.4$ in the reproduced field and we detected $339$ halos at $10^{13}$ \SI{}{M_{\odot}} $\leq$ $M_{200}$ $<$ $10^{14}$ \SI{}{M_{\odot}} and $14$ halos at $10^{14}$ $M_{\odot}$ $\leq$ $M_{200}$ when we adopt the same criterion in the COSMOS data. When we try the number of least connected redshift bins from $1$ to $5$, the use of more than $3$ connected bins produces the smallest fraction of separate structures that are misidentified as clusters.

%
\begin{figure}
    \centering
    \includegraphics[width = 0.45\textwidth]{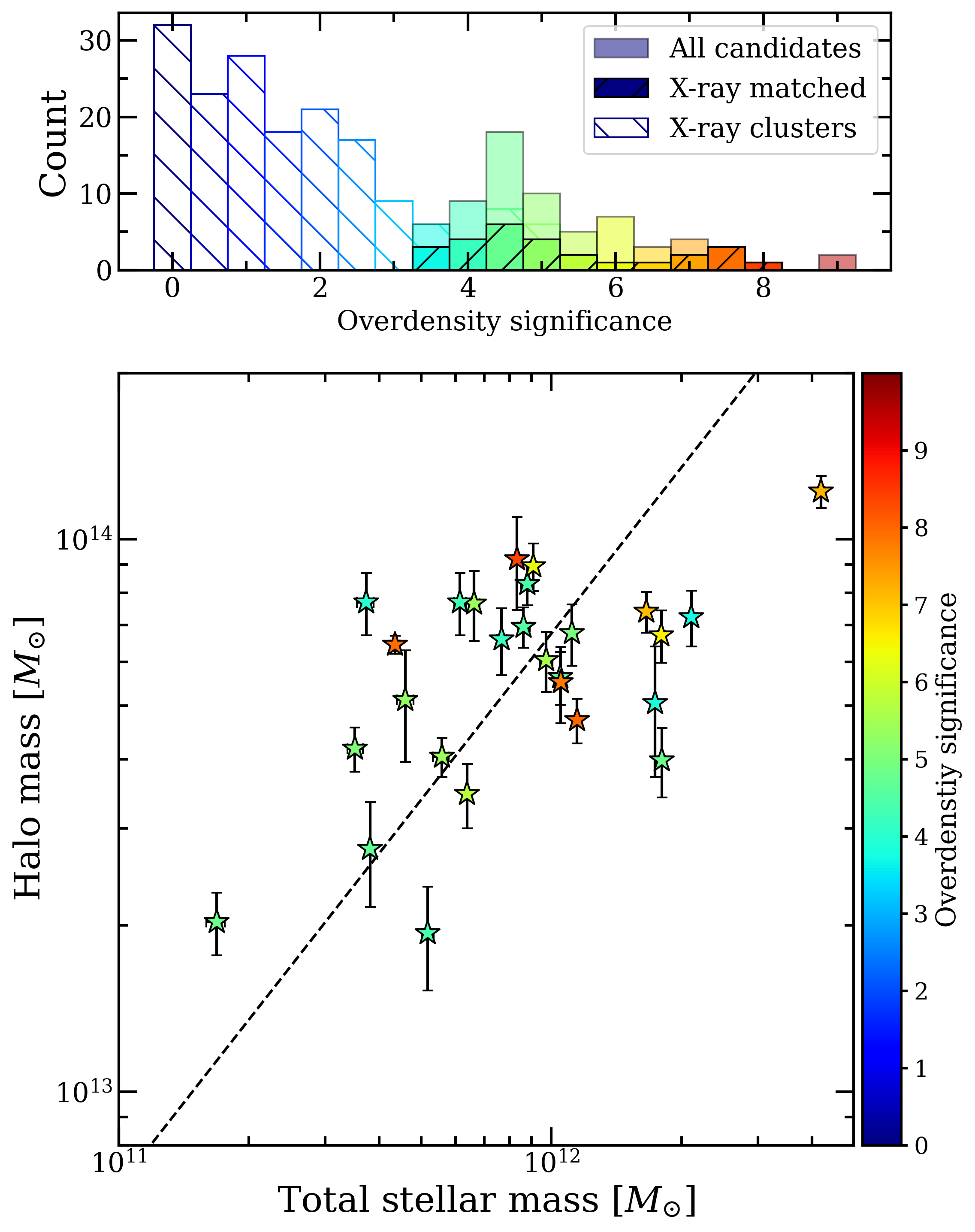}
    \caption{The comparison of cluster candidates found in this study and confirmed in the X-ray observation. The overdensity significance represents the number of standard deviations by which the density field deviates from its mean. Upper panel: The clusters that we identify have significant overdensities by definition while X-ray groups are more likely to be less dense and comprise a small number of member galaxies. Lower panel: The matched clusters show a statistical correlation (Pearson correlation coefficient = 0.55) between the total stellar mass of member galaxies and halo mass estimated from X-ray detected groups. The best-fit linear regression between the two masses among cluster candidates is displayed as a black dashed line.}
    \label{fig:mass_relation}
\end{figure}
In addition, $27$ out of $68$ cluster candidates are matched with the X-ray galaxy group catalog \citep{gozaliasl2019chandra}. The X-ray groups that are not identified in this study include only a small number of member galaxies. On the contrary, the cluster finding method based on overdensities cannot detect sparsely distributed members or a small number of members that have low overdensity significances as shown in the upper panel of Figure \ref{fig:mass_relation}. Since X-ray groups are known to be biased to a more dynamically relaxed system than optically selected groups \citep{2017MNRAS.472.1482O, 2021Univ....7..139L}, we can speculate that our samples include overdensities not fully collapsed. The X-ray groups that are not detected in our samples with significant overdensities are all located near the survey edges ($<$ \SI{1}{\mega pc}) and bright source masks. 

We estimate cluster halo masses ($M_{200}$) using the total stellar mass of member galaxies. To calibrate the mass estimator based on the total stellar mass, we compared the total stellar mass of member galaxies from this work to the X-ray-derived halo mass from \citet{gozaliasl2019chandra}. The X-ray halo masses in the \citet{gozaliasl2019chandra} data are derived from the X-ray luminosity-halo mass relation with weak-lensing calibration from \citealt{2010ApJ...709...97L}. Note that their mass to X-ray luminosity relation suggests a scatter in $\log{(M_{200})}$ of about 0.2 to 0.3. Here, the total stellar mass is defined as the sum of stellar masses above ${10}^{8.75}$ \SI{}{M_{\odot}}. Figure \ref{fig:mass_relation} compares the total stellar masses and the X-ray halo masses of clusters, showing a broad correlation between the two quantities. The fitting result between the X-ray halo mass and the stellar mass sum shows $M_{\textrm{200}} \propto\, 67.5^{+7.8}_{-7.0} \times$ total stellar mass. The derived halo masses are listed in Table \ref{tab:clusters}. We note that we will use the terms ``overdensity" and ``galaxy cluster" interchangeably for the cluster candidates.

\section{Results}

\subsection{Galaxy Evolution from Star-forming to Quiescent
Phase}\label{subsec:galaxy_evolution_from_sf_to_q}

A quiescent galaxy is defined as a galaxy that satisfies Eq. \ref{eq:quiescent_flag} where t(z) [$\textrm{yr}^{-1}$] is the age of the universe at redshift $z$ \citep{damen2008evolution, lee2015evolution}.

\begin{equation}
    sSFR < 1/3t(z)
    \label{eq:quiescent_flag}
\end{equation}

This definition takes into account the evolution of specific star formation rate (\textit{sSFR}) as a function of redshift and specifies quiescent galaxies as those that have relatively low \textit{sSFRs} at a given redshift. We adopt \textit{sSFR} values derived from SED fitting with \texttt{LePhare}.

Alternatively, it is also possible to use the color plane to select the passive galaxies. For example, quiescent galaxies can be identified by a two-color selection $UVJ$ method \citep{2005ApJ...624L..81L,2007ApJ...655...51W,williams2009detection}. COSMOS2015 \citep{laigle2016cosmos2015} and COSMOS2020 \citep{weaver2022cosmos2020} adopted the $NUV-r^{+}$ vs. $r^{+}-J$ criteria where quiescent galaxies meet the conditions $NUV - r^{+} > 3.1$ and NUV - $r^{+} > 3(r^{+} - J) + 1$. This method is known to avoid a mix between quiescent galaxies and dusty star-forming galaxies. However, some of the quiescent galaxies at higher redshifts are still misclassified to be star-forming galaxies because of uncertainties in their rest-frame colors \citep{weaver2022cosmos2020}. Since the classification from both color and $sSFR$ criteria exhibits nearly identical result, our study leans towards classifying galaxies based on the \textit{sSFR} for consistent comparison with \citetalias{lee2019more}. However, care should be taken with the \textit{sSFR} from SED fitting, given its systematic scatter and bias over time \citep{2015A&A...579A...2I, 2019MNRAS.486.5104L}. 

We investigate the difference between the results based on color and \textit{sSFR} selection. Among the $86,289$ galaxies more massive than $10^{8.99}$ \SI{}{M_{\odot}} at $0.1 \leq z \leq 1.4$, we find $14,052$ quiescent galaxies using the color selection and $17,777$ using the \textit{sSFR} selection. Notably, \SI{95.3}{\%} ($13,392/14,052$) of the quiescent galaxies identified with the color selection are also flagged as quiescent based on the \textit{sSFR} criterion in this study. The rest ($660/14,052$) are situated near the \textit{sSFR} selection cut. Similarly, galaxies categorized as quiescent only through \textit{sSFR} ($4,385/17,777$) are found located near the $NUV-{r}^{+}$ vs. ${r}^{+}-J$ color selection boundary. Of those, $3,561/4,385$ are residing in the color space of star-forming galaxies within 0.1 dex from the selection cut. We tried another selection criterion, $\log{(sSFR)} < \log{(sSFR_{\textrm{MS}}}) - 0.6$ \citep{2012ApJ...754L..29W}, where $sSFR_{\textrm{MS}}$ is the \textit{sSFR} [$\textrm{yr}^{-1}$] of the main sequence from \citet{2014ApJS..214...15S}. Only 19/17,777 quiescent galaxies based on the \textit{sSFR} selection are regarded as star-forming and 926/17,415 for vice versa, showing that our criterion is nearly identical to the criterion of $\log{(sSFR)} < \log{(sSFR_{\textrm{MS}})} - 0.6$.

In other words, most galaxies flagged as quiescent galaxies by only one of the selection methods are marginally missed by the other. The three selection methods select galaxies with very similar properties, with a slight difference in selection boundary. Therefore, we suggest our analysis is not sensitive to the selection method. We adopt the \textit{sSFR} method as justified above, and conducted the same analysis for the $NUV-r^{+}$ vs. $r^{+}-J$ and $sSFR < 10^{-11}\textrm{yr}$ galaxy classifications. The results are nearly identical, so we will present only the results based on the \textit{sSFR}-based galaxy classification. 

\begin{figure}
    \centering
    \includegraphics[width = 0.45\textwidth]{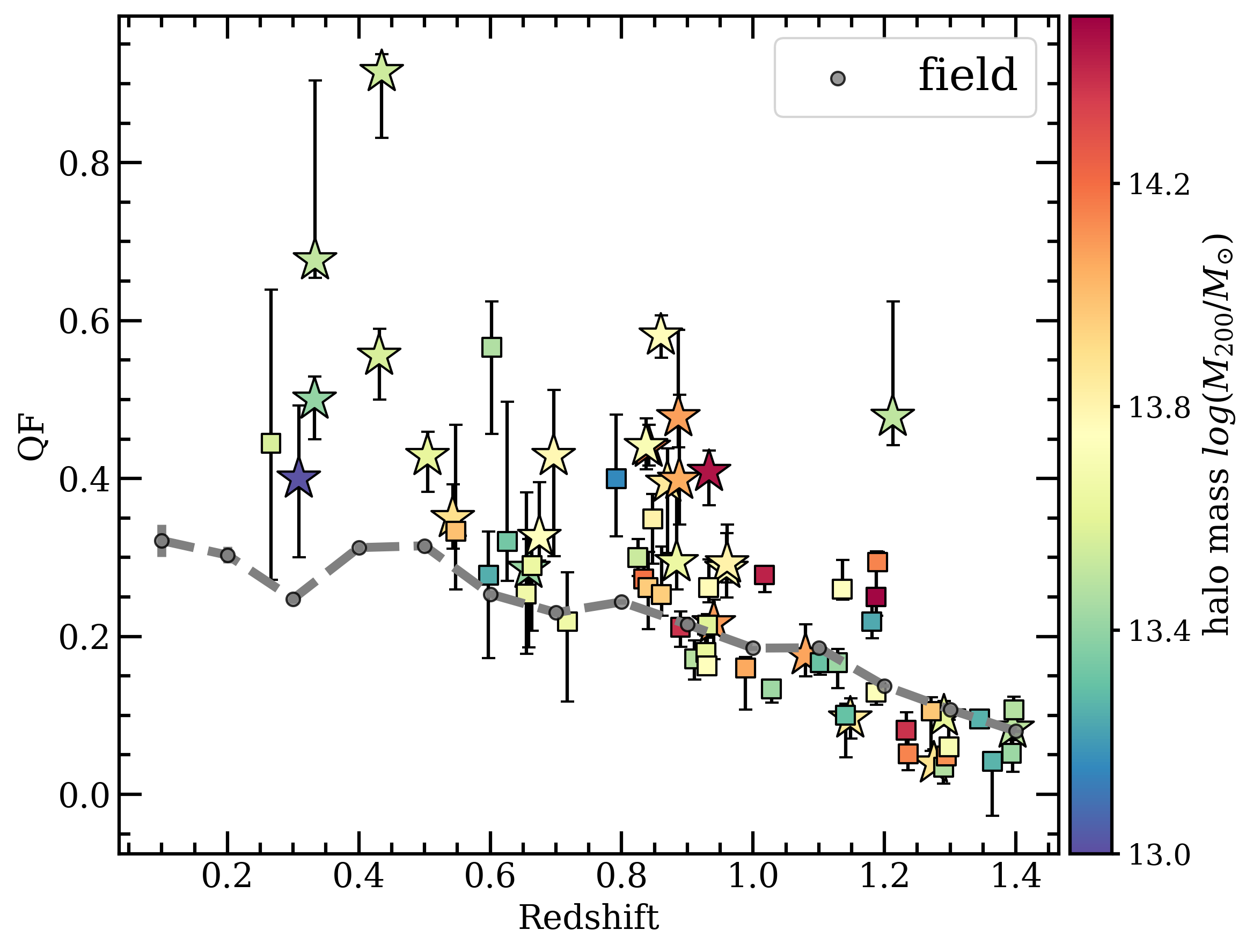}
    \caption{The distribution of quiescent galaxy fraction and redshift in the mass limited sample. The error bars represent the $68\%$ confidence interval, obtained from $1,000$ iterations of determining the membership by adding the error to the redshift center of the cluster. The errors follow a normal distribution $N(0, \sigma(1+z))$, where $\sigma(1+z)$ corresponds to the photometric redshift uncertainty at a given redshift. For comparison, the quiescent galaxy fraction from the field is overlaid with a dashed line.}
    \label{fig:qf_evolution}
\end{figure}

The quiescent galaxy fraction, hereafter abbreviated as \textit{QF}, denotes the number of quiescent galaxies over the total number of member galaxies. We use \textit{QF} as an indicator of star formation activity in galaxy clusters since other measures, such as the total or median star formation rate, can be easily biased by the amount of dust extinction, which is not well constrained without deep infrared data. Meanwhile, the fractional parameter \textit{QF} cancels out this effect and provides a more consistent metric regardless of the different assumptions involved in calculating SFR. Figure \ref{fig:qf_evolution} shows \textit{QFs} in galaxy clusters as a function of redshift. As the redshift increases, \textit{QF} decreases, consistent with the Butcher-Oemler effect \citep{butcher1978evolution}. The intuition of the web-feeding model can be found here from the distribution of varying \textit{QF}. At a given redshift and halo mass bin, \textit{QFs} of galaxy clusters have a wide range, which hints at the role of environment that influences the star formation activity or other physical parameter dependence.

\subsection{Reliability of 2D Density Field}

\begin{figure*}
    \centering
    \begin{tabular}{cccc}
    \includegraphics[width = 0.22\textwidth]{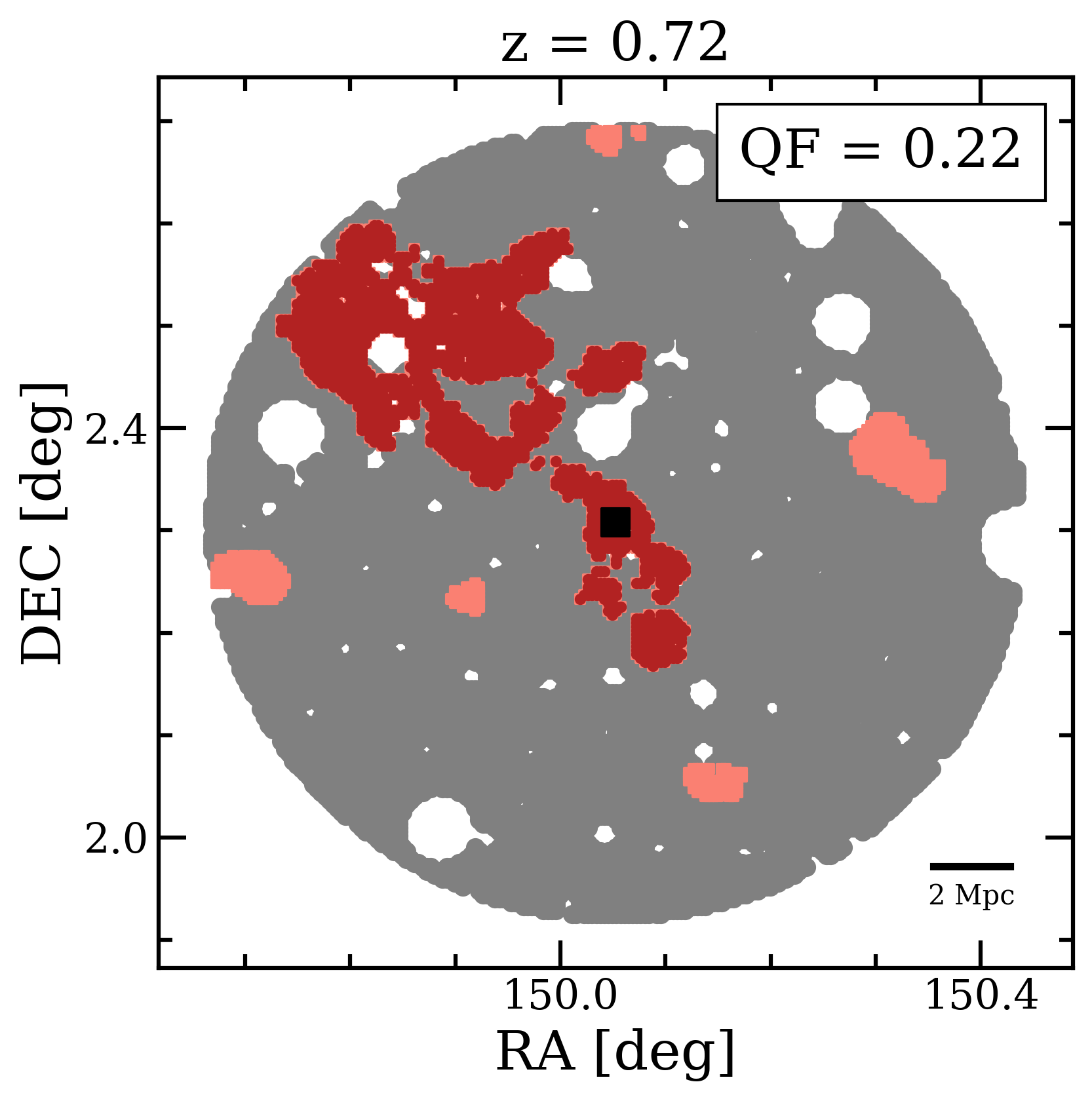} &
    \includegraphics[width = 0.22\textwidth]{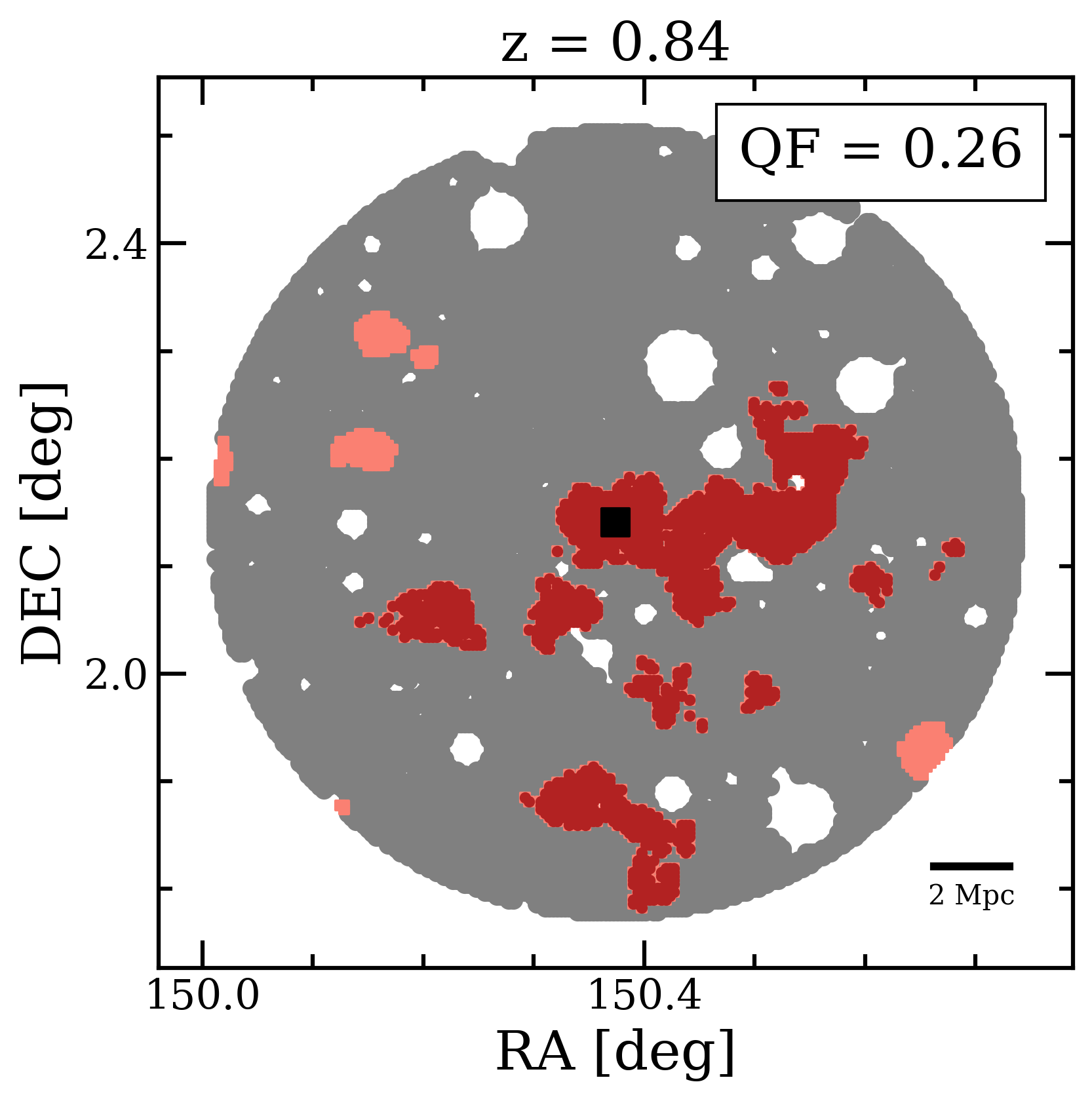} &
    \includegraphics[width = 0.22\textwidth]{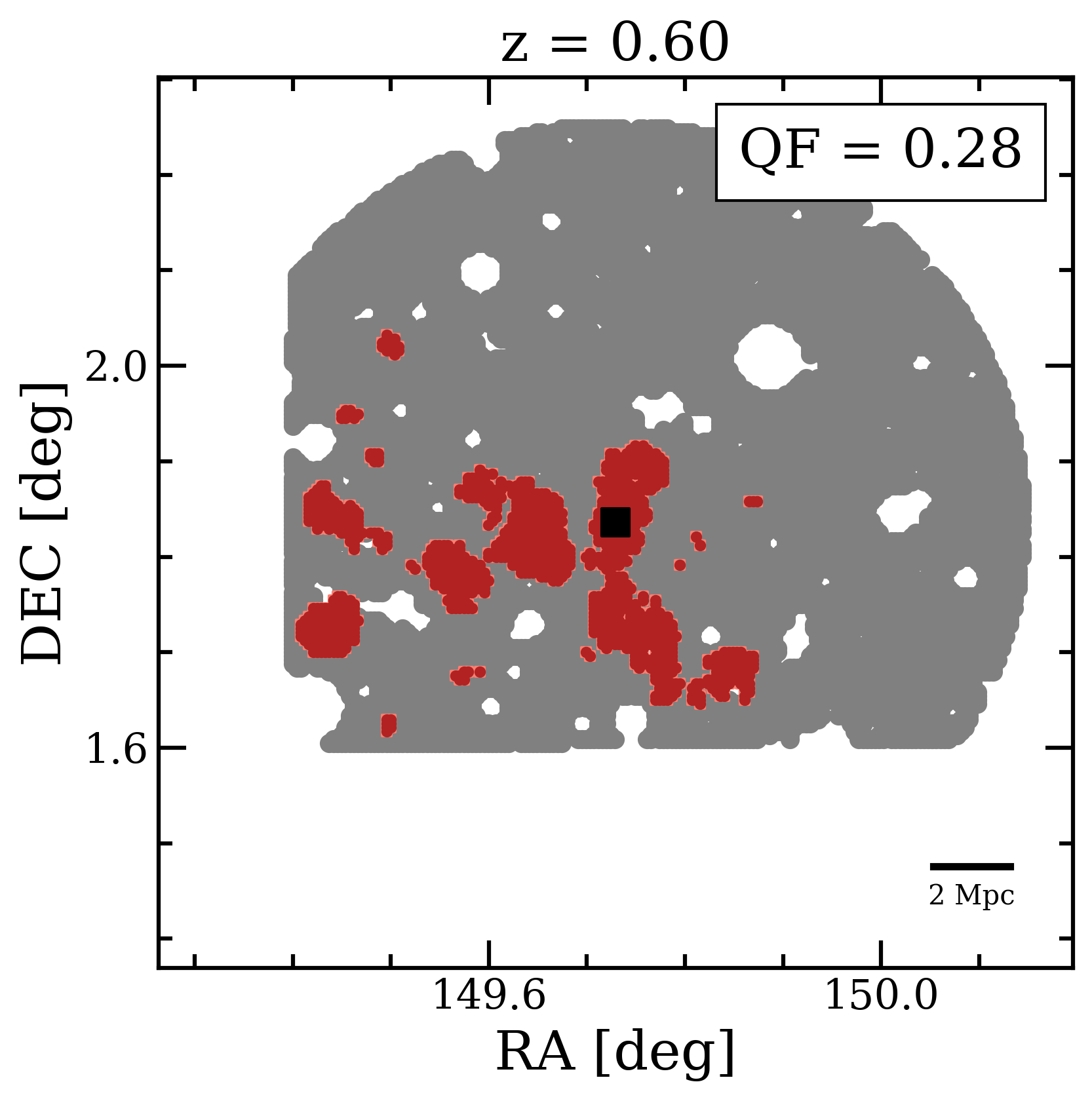} &
    \includegraphics[width = 0.22\textwidth]{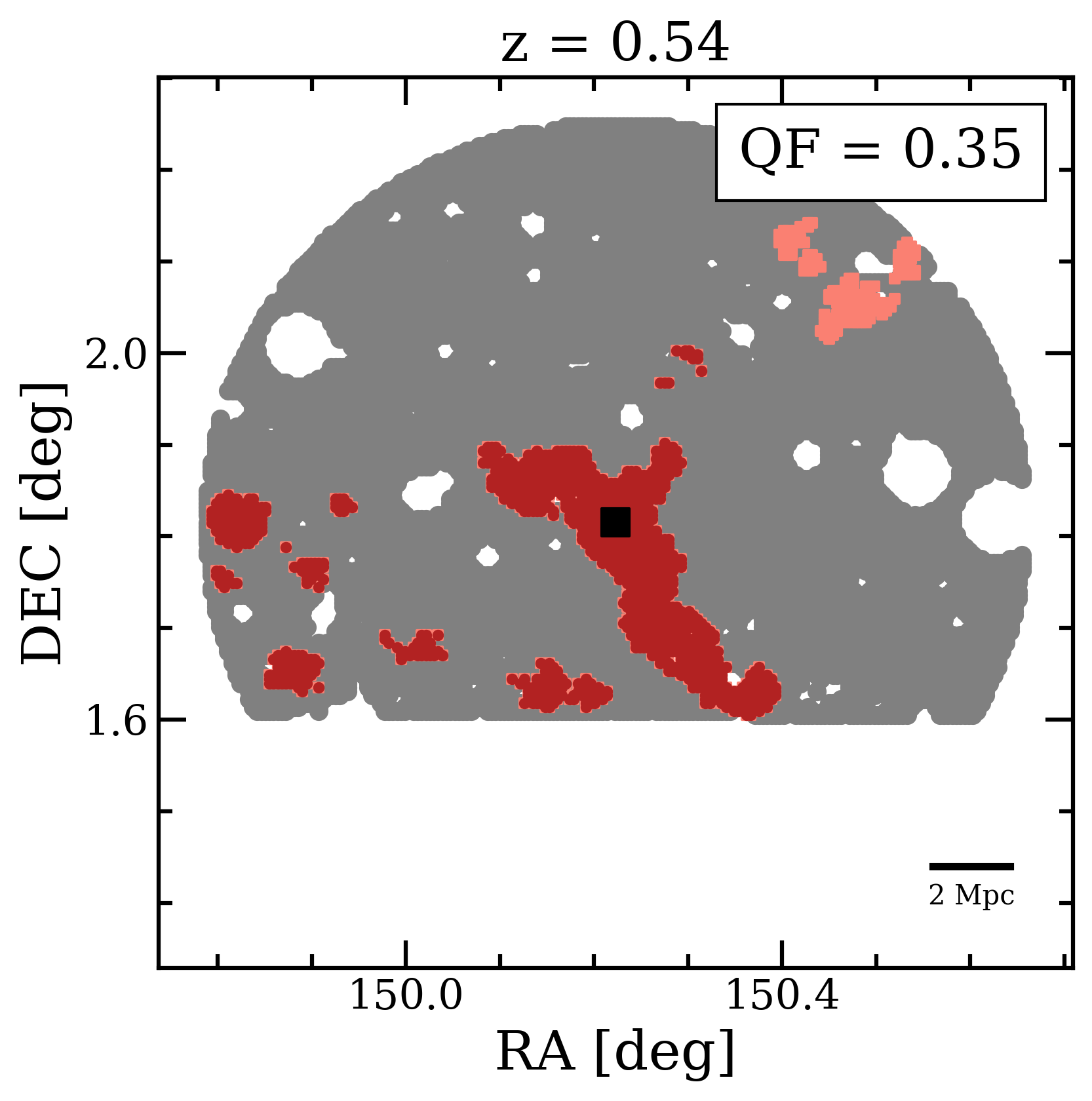} \\
    \includegraphics[width = 0.22\textwidth]{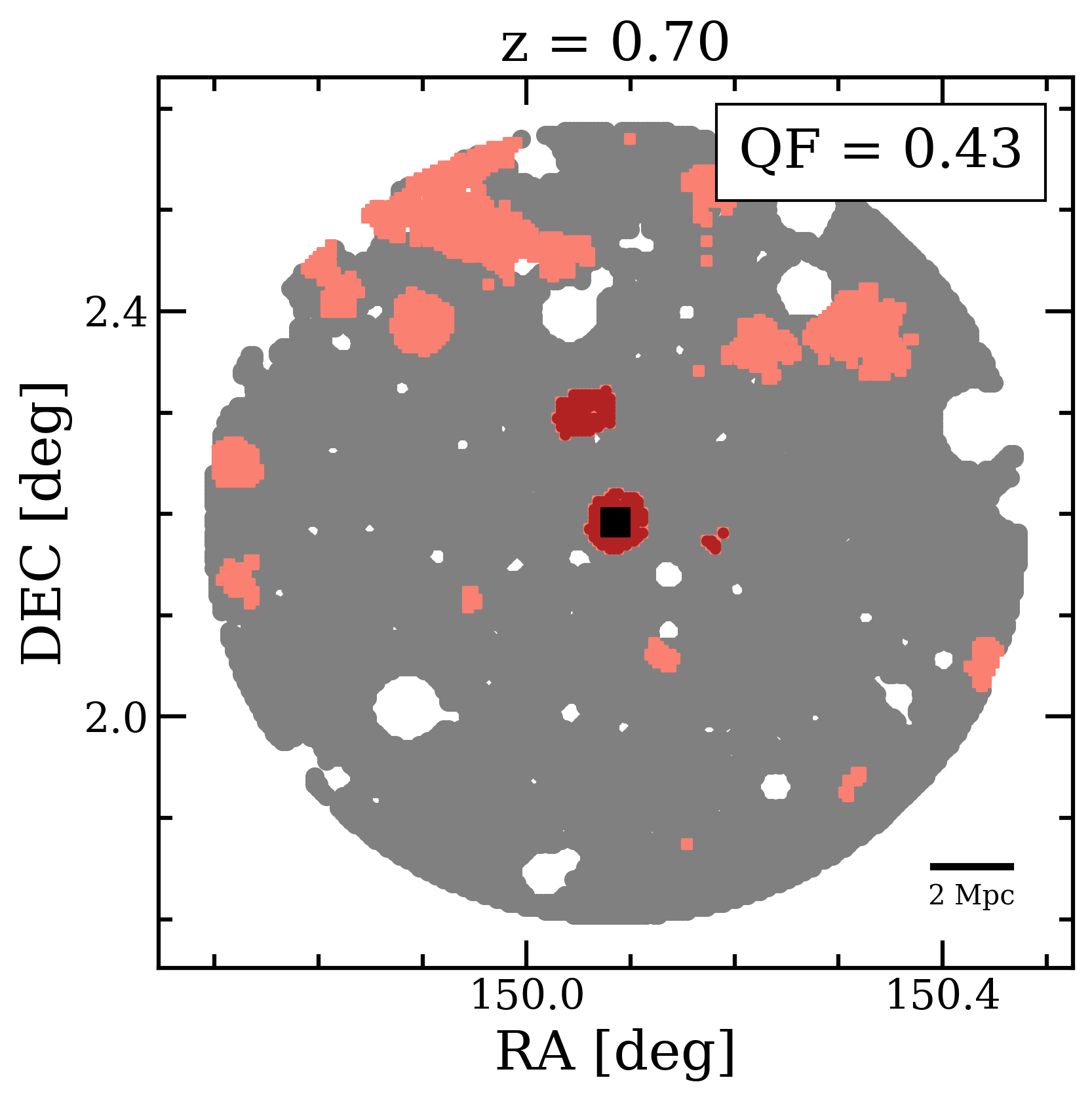} &
    \includegraphics[width = 0.22\textwidth]{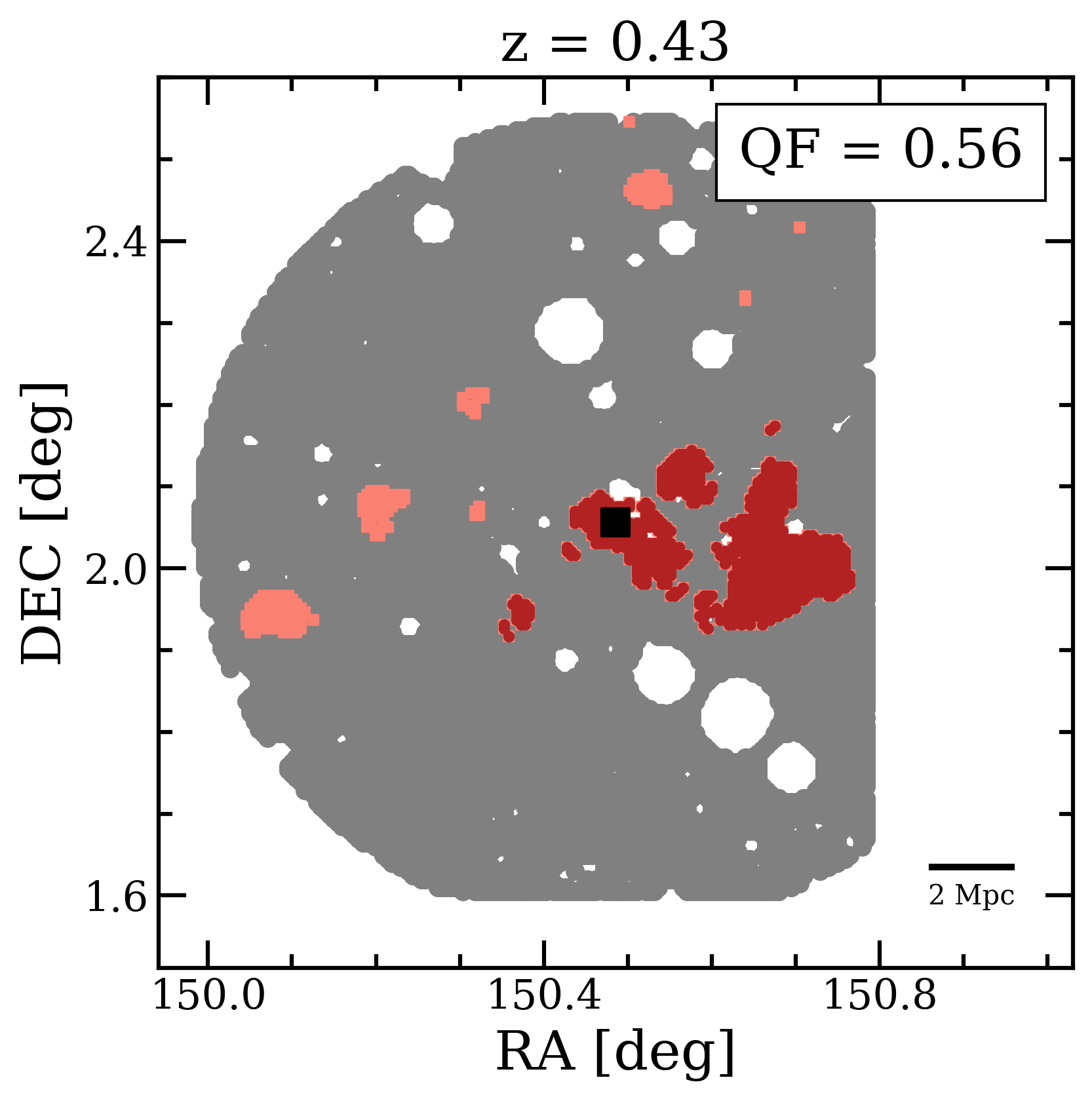} &
    \includegraphics[width = 0.22\textwidth]{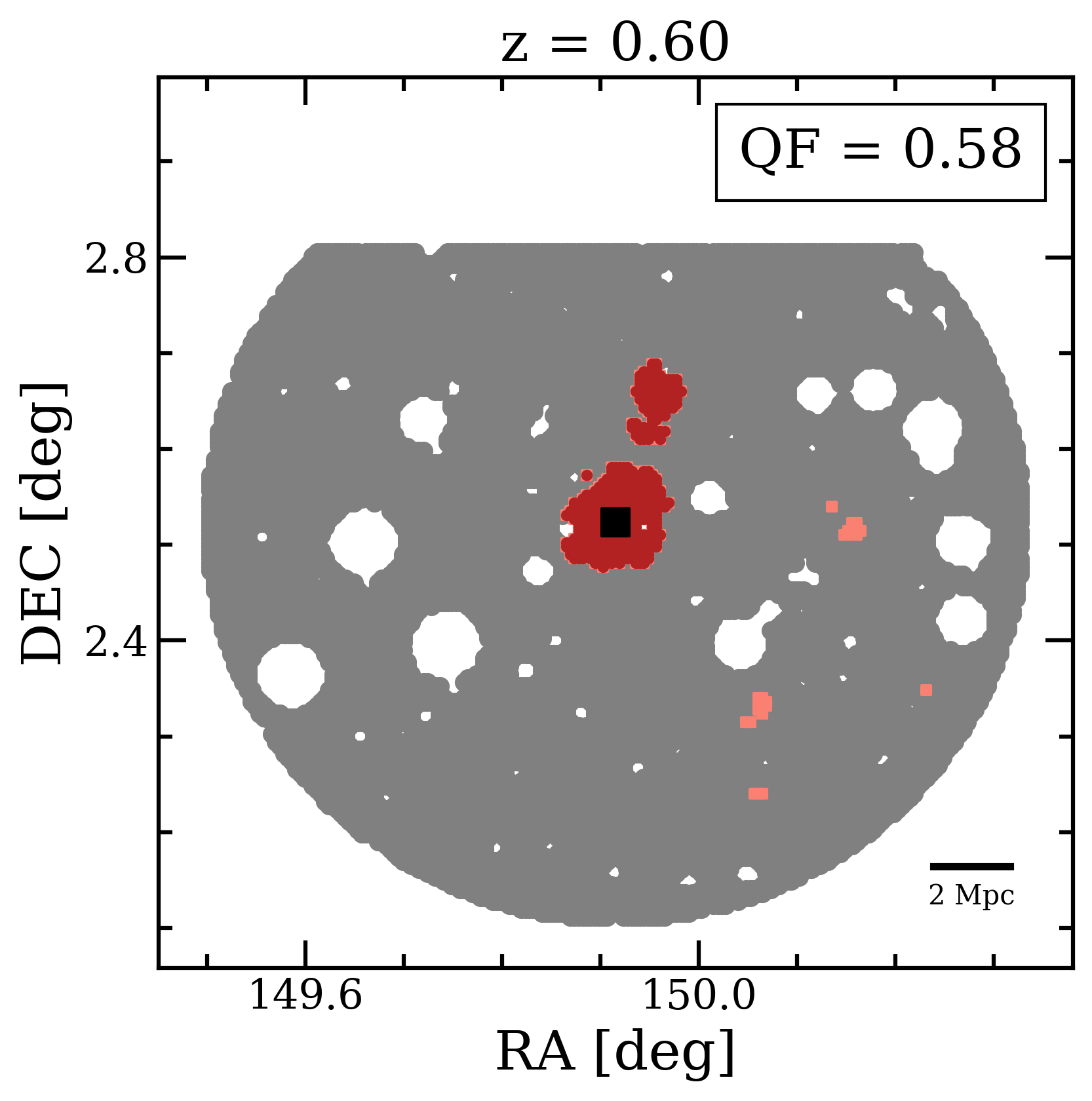} &
    \includegraphics[width = 0.22\textwidth]{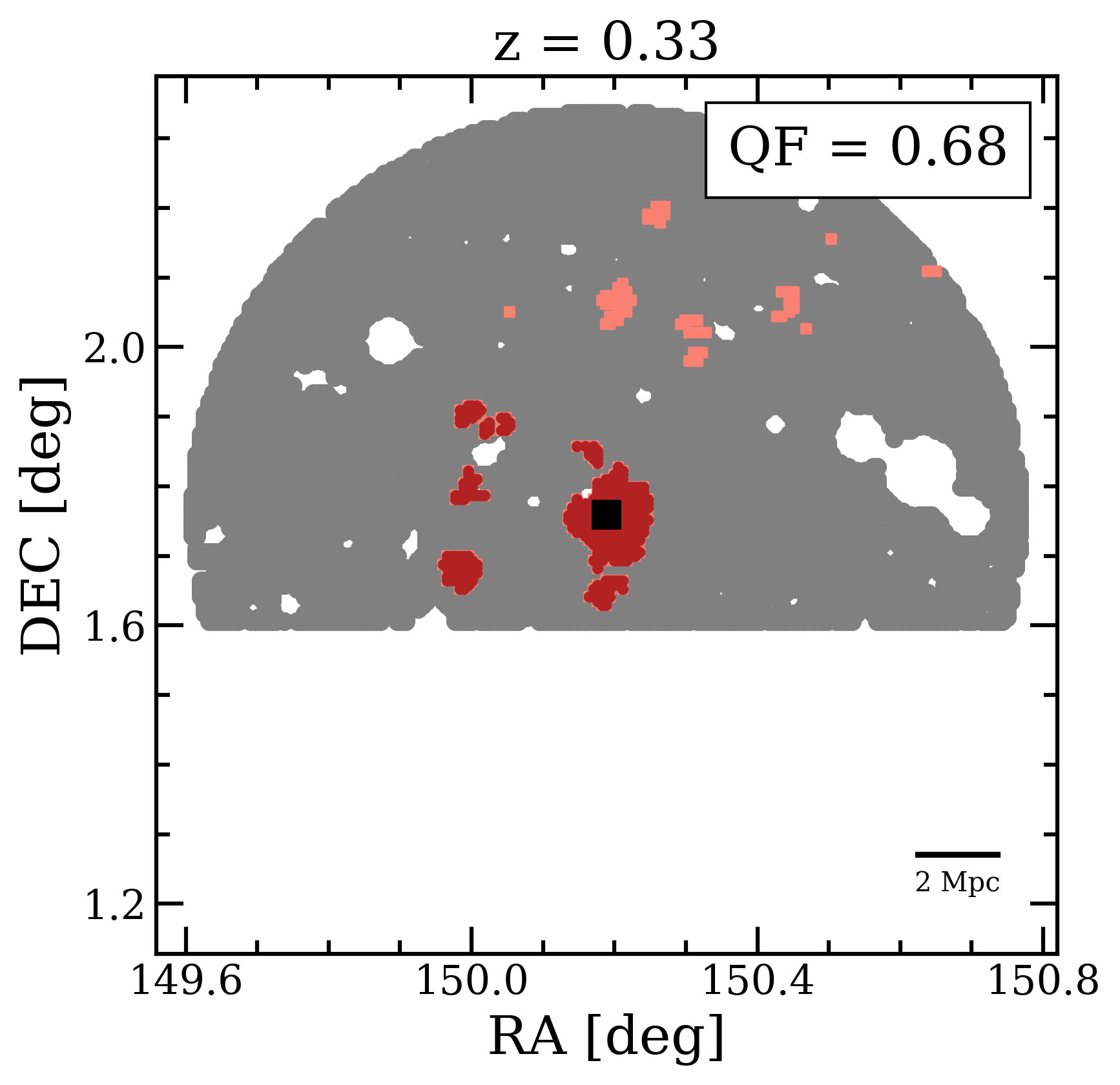}\\
    \end{tabular}

    \caption{Examples of overdense areas showing how the \textit{FoF fraction} is defined. Each panel displays overdensities (red region) connected to host cluster candidates (black square) in the increasing \textit{QF} order (left to right). The \textit{FoF Fraction} is defined as the ratio of the red area to the whole area (gray+red+pink areas). Unconnected overdensities within a \SI{2}{\mega pc} linking length are depicted in the pink region.}
    \label{fig:mcma_example}
\end{figure*}

As a quantitative proxy of the connected structure to a galaxy cluster, we define the term \textit{Friends-of-Friends fraction} (hereafter \textit{FoF fraction}) as the ratio between the total area of the $2\sigma$-level projected overdense regions connected with a 2 Mpc linking length by the Friends of Friends algorithm (\citealt{davis1985evolution}; red region in Figure \ref{fig:mcma_example}) and the projected area within a radius of 10 Mpc from the cluster (gray+red+pink region). In simple terms, this \textit{FoF fraction} characterizes the channel where large-scale cosmic web-feeding can take place. Because we are interested in the inter-cluster scale, we restrict our analysis to the environment within 10 Mpc. Additionally, we use 2 Mpc as the linking length, which aligns with the typical size of galaxy clusters and is short enough to account for interactions among galaxies or groups. We checked that the variation of linking length (0.5, 1.0, 1.5, 2.0 Mpc) does not significantly change the overall results.

We note that there have been various methods to measure the large-scale cosmic web (e.g., \citealt{sousbie2011persistent,2013MNRAS.429.1286C,2014MNRAS.438..177A, 2014MNRAS.438.3465T,2018MNRAS.473.1195L}). Admittedly, there may be better ways to analyze the effect of the web-feeding model than the \textit{FoF fraction}. However, we decided to adopt the \textit{FoF fraction} for comparing our results with \citetalias{lee2019more} in a consistent way by using the same metric. Since defining the large scales is subject to the choice of the measurement method, uniformly gauging the impact of the scales of our interest (inter-cluster $\sim$Mpc) is challenging. With a cosmological simulation, we confirm that galaxies and small groups are infalling following the 2$\sigma$ overdensities connected to the host cluster in Section \ref{subsec:what_fuels_the_galaxy_cluster}. In the future, we hope to explore if there are better ways to calculate the web-feeding trend.

To sum up, the \textit{FoF fraction} indicates the volume (area) of the reservoir from which infalling galaxies, groups, or cold gas, if exists, originate. We refer to these infalling components as \emph{infallers} and expect them to impact the \textit{QF}. The precise influence of infalling galaxies and cold gas on the increase in star-forming galaxies in clusters is not clear. We will discuss the role of gas on cluster galaxies in Section \ref{subsec:what_fuels_the_galaxy_cluster}. Therefore, we refer to all the different ingredients fueling a cluster to keep \textit{QF} at a low value as infallers for simplicity.

\begin{figure}
    \centering
    \includegraphics[width = 0.45\textwidth]{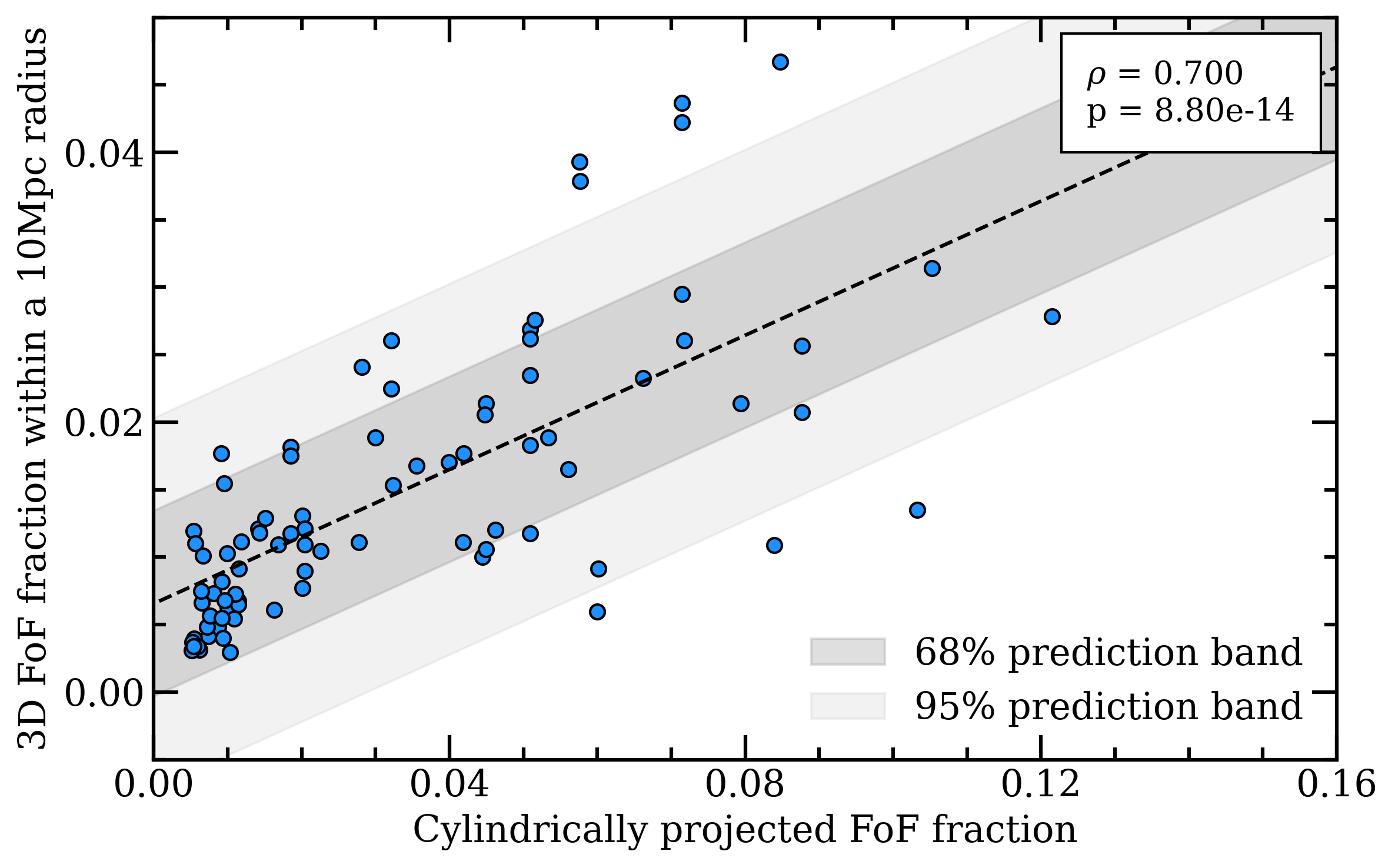}
    \caption{The $x$-axis denotes the 2D projected cylindrical \textit{FoF fraction} in the lightcone mock catalog \citep{merson2013lightcone} derived with the same method for COSMOS2020. On the other hand, the $y$-axis denotes the 3D spherical \textit{FoF fraction}, taking into account a physical distance of \SI{10}{\mega pc} in the same mock data. They exhibit a general correlation within the $95\%$ prediction level. The Pearson correlation coefficient is $\rho$ and p-value $p$. The best-fit linear regression line is shown as a dashed line.}
    \label{fig:mock_FoF}
\end{figure}

Before testing the web-feeding model, we check if the 2D projected structures can represent actual 3D structures. Using the same galaxy light-cone mock catalog \citep{merson2013lightcone} employed to verify the cluster-finding method, we calculate the relationship between the \textit{FoF fraction} derived from (1) a cylindrical region, with a projected physical radius of \SI{10}{\mega pc} and a height corresponding to the photometric redshift uncertainty $0.01(1+z)$ and (2) a spherical region within a physical radius of \SI{10}{\mega pc} from the cluster center. Figure \ref{fig:mock_FoF} shows a moderate correlation between the 2D and 3D \textit{FoF fractions} with a correlation coefficient of $0.700$. Several previous studies (e.g., \citealt{darvish2017cosmic, laigle2018cosmos2015}) have also demonstrated that 3D cosmic web can be reliably traced from 2D counterparts up to $z \sim 1$ with a photometric redshift uncertainty of the order of $0.01(1+z)$.

\subsection{Web-Feeding in the COSMOS Field}
 
\begin{figure}
    \centering
    \includegraphics[width = 0.45\textwidth]{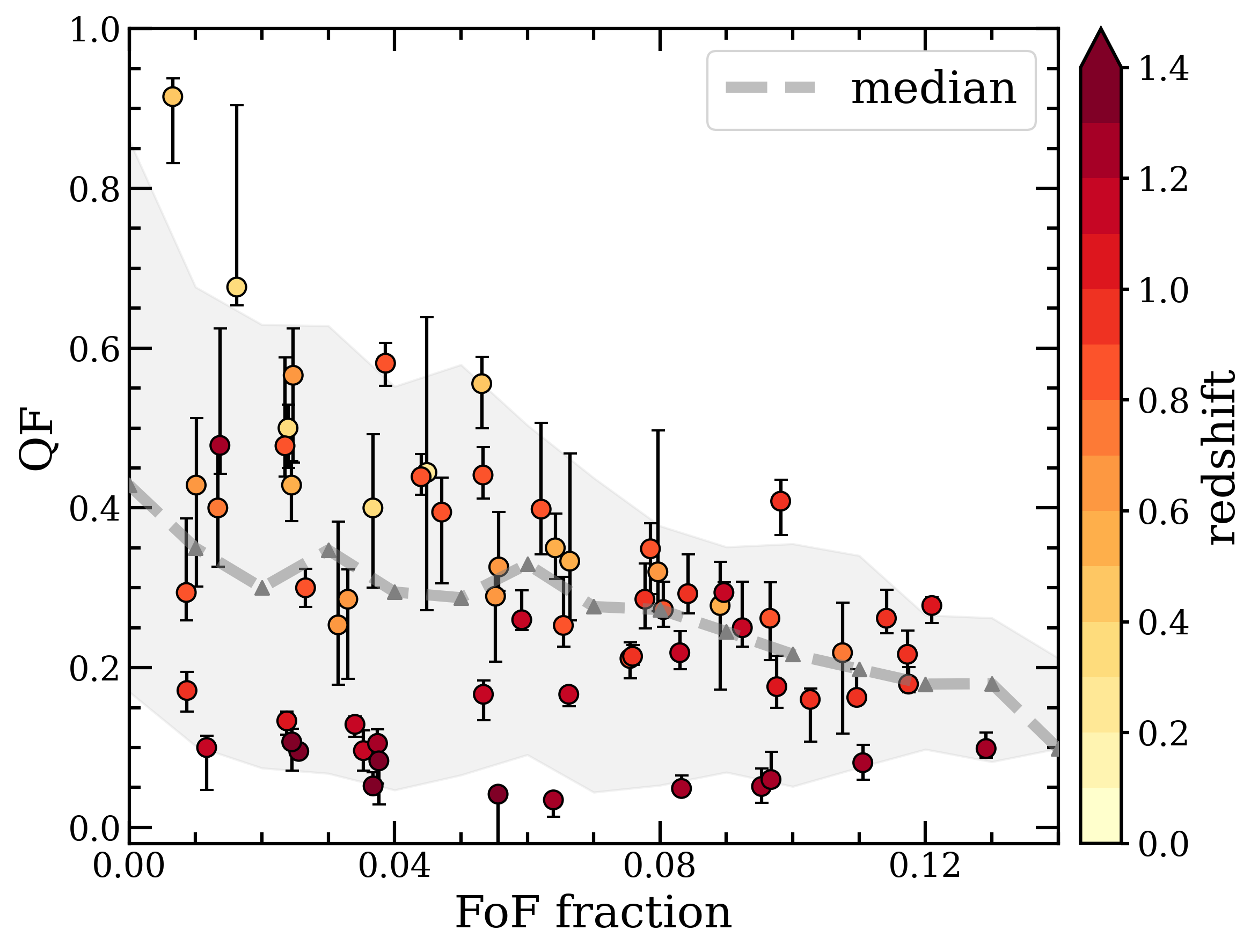}
    \caption{\textit{QF} as a function of \textit{FoF fraction} for the total $68$ galaxy clusters found in the COSMOS field. The color code means the redshift of a given cluster. The grey dashed line is plotted as median \textit{QF} at given \textit{FoF fraction} with $1\sigma$ confidence level (grey shade).}
    \label{fig:WFM}
\end{figure}

Figure \ref{fig:WFM} shows the relationship between \textit{FoF fraction} and \textit{QF} covering the overall redshift range ($0.1 \leq z \leq 1.4$). The lower \textit{FoF fractions} exhibit a broad range of \textit{QFs} whereas higher \textit{FoF fractions} are mostly associated with low \textit{QFs}. The Pearson correlation coefficient for \textit{FoF fraction} and \textit{QF} is $-0.401$. Even though the correlation itself is not strong, it remains significant given the p-value ($0.0007$) and the general trend is consistent with the result of \citetalias{lee2019more}. This trend is visually demonstrated in Figure \ref{fig:mcma_example}, where we show the density map with the cluster \textit{QF} and the connected LSS are indicated. In the upper panels, clusters with lower \textit{QFs} are shown to have LSSs connected to them (i.e., higher \textit{FoF fraction}). Similarly, clusters with higher \textit{QFs} are found to be relatively isolated (i.e., lower \textit{FoF fraction}) in the lower panels. This supports the web-feeding effect of the large-scale cosmic web on star formation activity in galaxy clusters. Therefore, galaxy clusters with low \textit{QF} generally populate largely connected environments rather than isolated areas. 

\subsubsection{The Effect of Redshift and Halo Mass}

\begin{figure*}
    \centering
    \includegraphics[width = 0.95\textwidth]{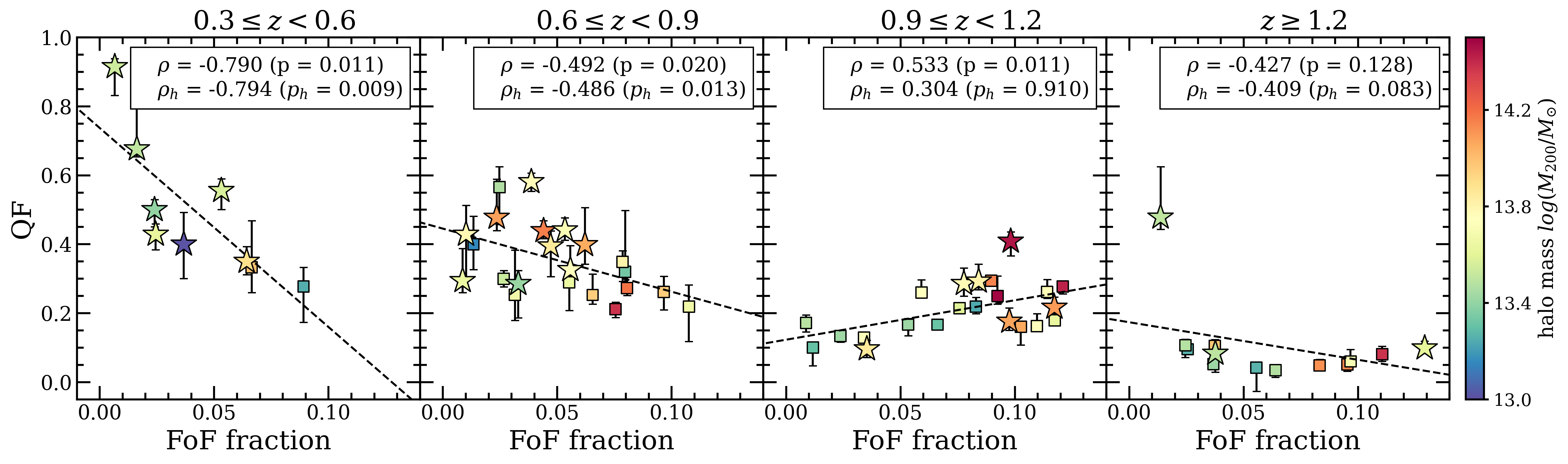}
    \caption{The anti-correlation between \textit{QF} and \textit{FoF fraction} in 4 redshift bins. The Pearson correlation coefficient and its associated p-value are denoted as $\rho$ and $p_h$, respectively, where the subscript $h$ signifies fixed halo masses. The halo mass is represented as a color of marker. The star-shaped markers correspond to the clusters that have been detected in the X-ray group catalog \citep{gozaliasl2019chandra} while those in square shape are candidates found based on photometric redshifts in this study.}
    \label{fig:WFM_zbin}
\end{figure*}

While these results are consistent with the web-feeding model, it is well known that \textit{QF} is also dependent on the cluster halo mass and environment (e.g., \citealt{2012MNRAS.424..232W}). In order to separate the effect of redshift evolution and halo mass, we divided the redshift bins into 4 intervals ($0.3 \leq z < 0.6$, $0.6 \leq z < 0.9$, $0.9 \leq z < 1.2$, $z \geq 1.2$) as shown in Figure \ref{fig:WFM_zbin}. For the two lower redshift bins at $0.3 \leq z < 0.6$ and $0.6 \leq z < 0.9$, the correlation between \textit{FoF fraction} and \textit{QF} is more pronounced than the whole sample, with the correlation coefficients $\rho$ of $-0.790$ and $-0.492$ respectively. We will discuss the opposite trend at $0.6 \leq z < 0.9$ by constraining the halo mass. On the other hand, no clear \textit{FoF fraction} dependence on \textit{QF} appears for higher redshift bin ($z \geq 1.2$). At this epoch, the growth of the overdensities is not as advanced as in those at lower redshifts where the trend of web feeding appears clearly. The other explanation is that most galaxies at high redshifts are not quenched yet unlike their counterparts at lower redshifts. In the earlier universe, the star formation activity in cluster members is still comparable to that of field galaxies (e.g., \citealt{brodwin2013era}), demonstrating that the correlation between \textit{QF} and \textit{FoF fraction} does not stand out. 

Following the nature of the web-feeding model, the accretion of galaxies is more likely to be strong at the site where the gravitational potential is the deepest. For the lower redshift bins at $0.3 \leq z < 0.6$ and $0.6 \leq z < 0.9$, we also examined how the \textit{FoF fraction} vs. \textit{QF} trend changes depending on the $M_{200}$ values. The partial correlation coefficients $\rho_{h}$ when fixing halo mass at a given redshift bin are $-0.794$ (p-value = $0.009$) and $-0.486$ (p-value = $0.013$), showing nearly identical correlation. The result suggests that the \textit{FoF fraction} vs. \textit{QF} correlation exists independent of the $M_{200}$ dependence. While clusters lying at low redshift still follow the persistent relation with fixed halo masses, high redshift clusters still do not show any such trend. At $0.9 \leq z < 1.2$, the statistical analysis indicates that \textit{QF} is not related to \textit{FoF fraction} when considering fixed halo mass, although a correlation is observed when halo mass is not constrained. Consistently, the anti-correlation trend becomes insignificant at higher redshift $z \geq 1.2$. 

The observed web-feeding trend appears to diverge from the previous findings presented in \citet{darragh2019group} and \citet{kraljic2020impact}, where central galaxies in groups or clusters connected to more filaments (with higher connectivity, indicative of the large-scale cosmic web) \citep{2018MNRAS.479..973C}, are found to be less star-forming. We confirm that passive central galaxies in our cluster candidates do not show larger \textit{FoF fractions} than star-forming ones as suggested in \citet{darragh2019group}. \citet{darragh2019group}, relying on the \textsc{Horizon-AGN} simulation \citep{2014MNRAS.444.1453D}, speculated that groups with higher connectivity are more likely to have experienced a group major mergers in their past, which would have increased the connectivity (see also \citealt{2021A&A...651A..56G}), and, \textit{in the long term}, quenched the central galaxy due to the activity of the central AGN \citep{2005Natur.433..604D, 2016MNRAS.463.3948D}. We note however that the X-ray selection might be biased towards relaxed groups (e.g., \citealt{2017MNRAS.472.1482O, 2021Univ....7..139L, 2022A&A...665A..78S}) and concentrated structures containing an AGN (e.g., \citealt{2007ApJ...654L.115S, 2014ApJ...790...43O}). In this sense, the COSMOS X-ray group sample might miss those groups/clusters which are either structures not fully collapsed, like proto-clusters, or clusters specifically in the process of merging, and therefore containing galaxies with temporarily boosted star-formation (e.g., \citealt{2007A&A...468...61D, 2017MNRAS.472L..50M}). Indeed only 10 of our clusters overlap with groups in \citet{darragh2019group}. However, it is crucial to note that our investigations focus on the influence of the cosmic web extending beyond overdensity-based clusters ($\sim$\SI{10}{\mega pc}) on member galaxies. It remains plausible that satellite galaxies maintain their star formation within relatively dense environments, while massive central galaxies are more prone to quenching \citep{2012MNRAS.424..232W,2015MNRAS.450..435S, 2017MNRAS.472.3570H, 2022MNRAS.510..674W}. 

\begin{figure*}
    \centering
    \gridline{\fig{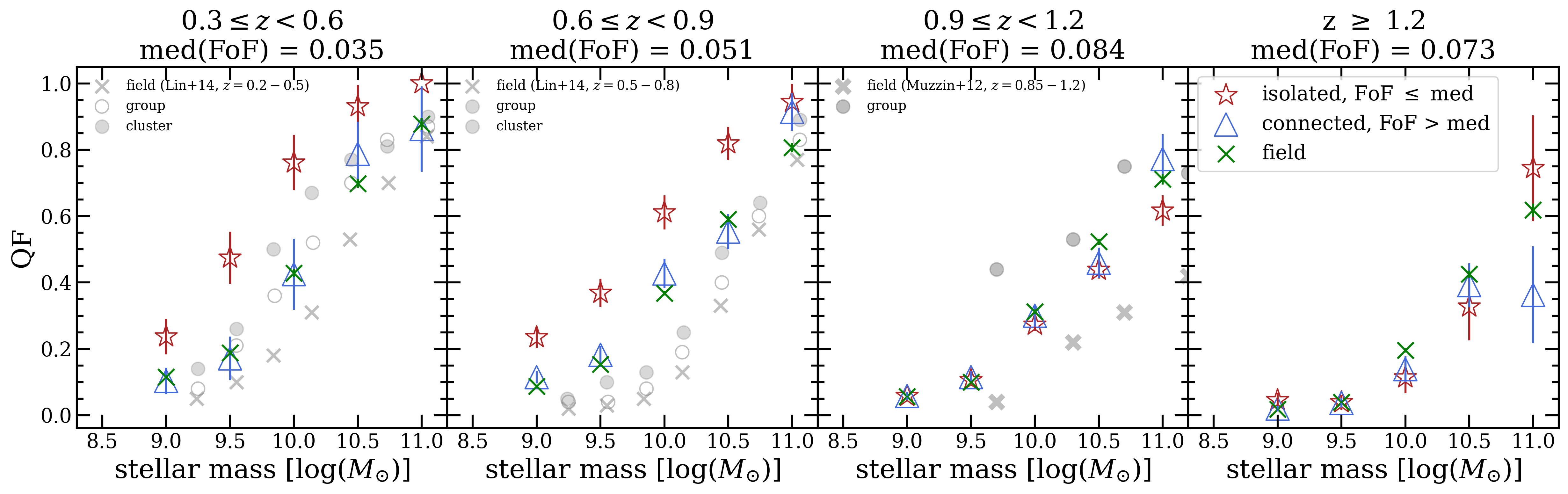}{0.95\textwidth}{(a)}}
    \gridline{\fig{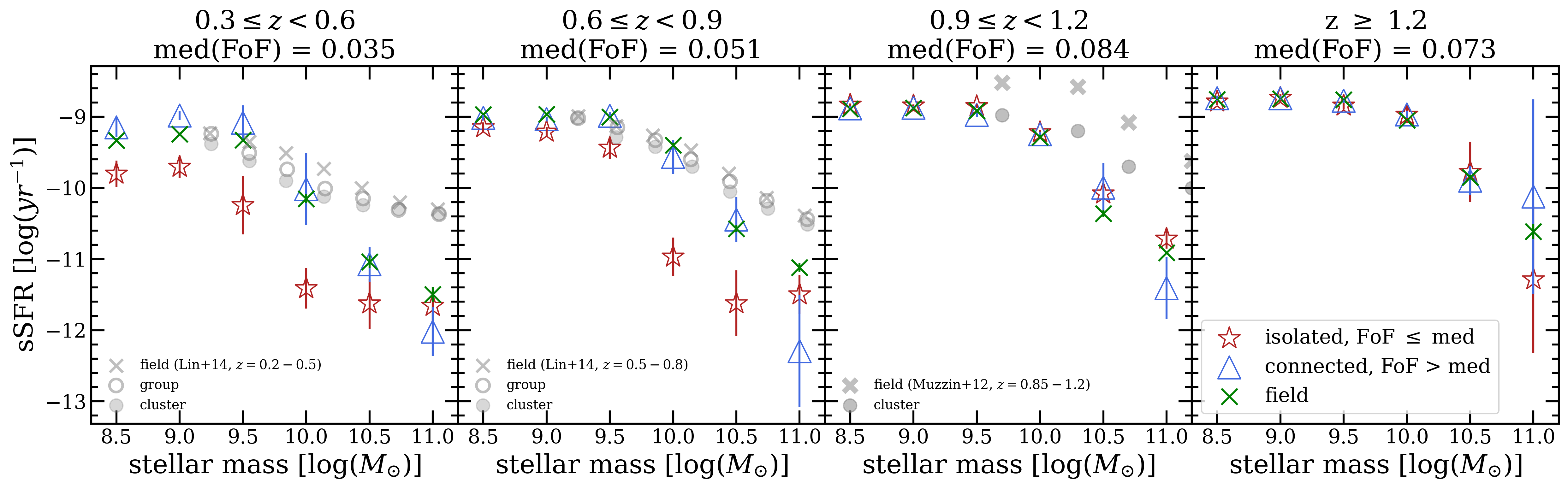}{0.95\textwidth}{(b)}}

    \caption{The median \textit{QF} (upper panel) and median \textit{sSFR} (lower panel) in each stellar mass bins of member galaxies. The member galaxies in clusters with \textit{FoF fractions} larger than the median at a given redshift bin (red star) are more actively forming stars than those in clusters with lower \textit{FoF fractions} (blue triangle). The green cross represents the case of field galaxies that are residing in the area with density $\leq 2\sigma$ for reference. The observational data from other literature \citep{2014ApJ...782...33L, 2012ApJ...746..188M} are overplotted in grey points. Here, we only compare the face values of \textit{sSFR} and \textit{QF} to see if their general trends with regard to stellar masses are consistent. Note that the criteria of quiescent/star-forming galaxies, environment (field, group, and cluster), and IMF models are different between studies.}
    \label{fig:stellarmass_in_obs}
\end{figure*}

\subsubsection{The Effect of Stellar Mass}

Since \textit{QF} is also dependent on the stellar mass of galaxies with \textit{QF} being higher for higher $M_{\ast}$ galaxies (e.g., \citealt{peng2010mass, 2013ApJS..206....3S, lee2015evolution}), we look into the \textit{QF} vs. \textit{FoF fraction} correlation further to see how the $M_{\ast}$ dependence plays out in the correlation. To accomplish this, we examine the \textit{QF} versus \textit{FoF fraction} trend by dividing the member galaxy sample by their $M_{\ast}$. Figure \ref{fig:stellarmass_in_obs} presents the median \textit{QF} and \textit{sSFR} for member and field galaxies in stellar mass bins at a given redshift bin, comparing those in \emph{connected} clusters, \emph{isolated"} clusters, and in \emph{field}. Here, the ``connected" clusters are defined as those with \textit{FoF fraction} larger than the median in the corresponding redshift bin, ``isolated" as those with \textit{FoF fraction} less than the median, and ``field" as those that do not belong to clusters or $2\sigma$ overdensities. 

The upper panels of Figure \ref{fig:stellarmass_in_obs} illustrate that the \textit{QFs} of isolated clusters are generally higher across most $M_{\ast}$ values than those of connected clusters at $z < 0.9$. Similarly, the \textit{sSFRs} tend to have lower values for isolated clusters compared to connected clusters (the lower panel of Figure \ref{fig:stellarmass_in_obs}). The connected clusters have \textit{QFs} and \textit{sSFRs} similar to galaxies in the field. However, beyond $z \geq 0.9$, the \textit{QF} distribution between the field, isolated, and connected clusters disappears. A similar trend is found for \textit{sSFR} of member galaxies. These results confirm the correlation between \textit{QF} and \textit{FoF fraction} at $z \lesssim 0.9$ regardless of $M_{\ast}$, as expected from the web-feeding model.

Notably, field galaxies show star formation activities similar to galaxies in connected clusters. The similarity in \textit{QF} or \textit{sSFR} between field galaxies and connected clusters reflects the influence of infalling galaxies keeping \textit{QF} relatively low. Such galaxies would be eventually quenched. A similar trend can be found for cluster and field galaxies studied in \citealt{2014ApJ...782...33L}.

\subsubsection{Concentration Parameter}

\begin{figure*}
    \centering
    \includegraphics[width = 0.95\textwidth]{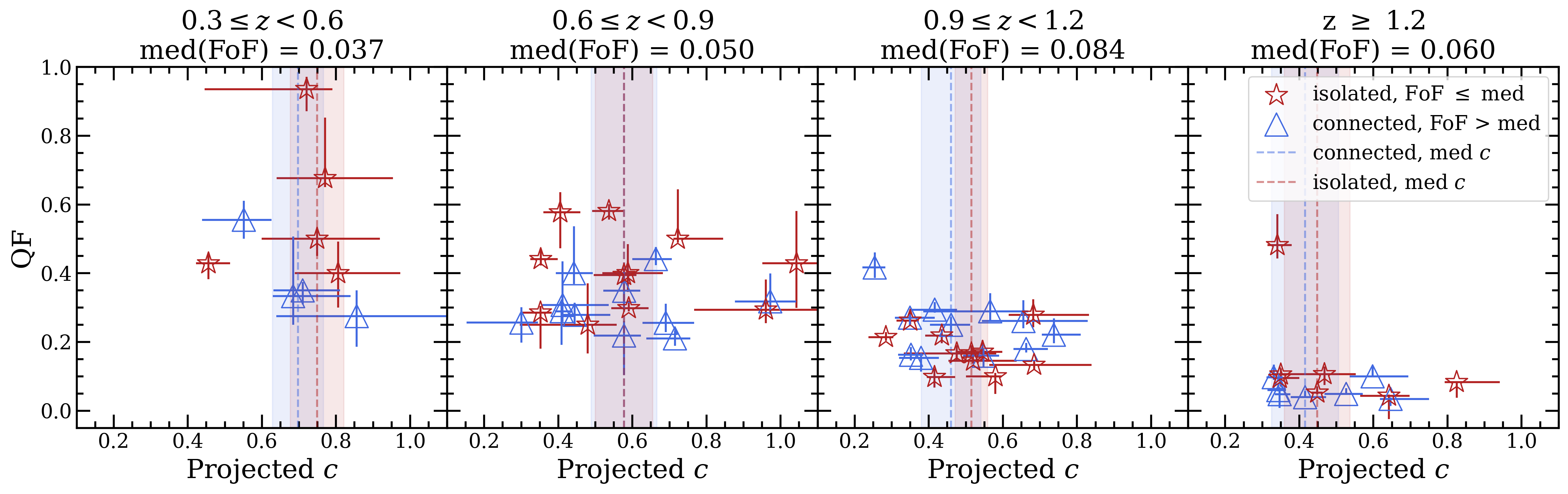}
    \caption{The comparison of concentration parameters and \textit{QFs} in connected (the blue triangles) and isolated clusters (the red stars). No notable difference for \textit{c} is found between the isolated and connected clusters.}
    \label{fig:spatial_distribution}
\end{figure*}

The physical difference between connected and isolated clusters is also investigated with projected concentration parameters \textit{c} defined as the ratio of the area where \SI{30}{\%} and \SI{70}{\%} of members reside. The projected concentration parameter serves as a practical proxy of the concentration parameter from the Navarro-Frenk-White (NFW) density profile \citep{navarro1997universal} when only photometric redshifts are available. In Figure \ref{fig:spatial_distribution}, we calculate the projected concentration parameters for each redshift bin and present those values in Table \ref{tab:projected_cpars}. 

Across all four redshift bins, no significant difference between the projected concentration parameters in connected and isolated clusters is observed. This tendency is consistent with the findings of \citetalias{lee2019more}, where the correlation between \textit{QFs} and \textit{c} is weak. If the clusters with high \textit{FoF fractions} are contaminated more by surrounding density structures, we expect to find a difference in the concentration parameter as a function of \textit{FoF fraction} values. No strong correlation with \textit{c} and \textit{FoF fraction} in Figure \ref{fig:spatial_distribution} and Table \ref{tab:projected_cpars} assures that the cluster selection is not biased by the surrounding structures. 

\begin{deluxetable}{ccc}
\tabletypesize{\scriptsize}
\tablewidth{0pt}
\tablecaption{The median projected concentration parameters and $1\sigma$ confidence interval with various \textit{FoF fractions} and redshifts.\label{tab:projected_cpars}}

\tablehead{\colhead{} & \colhead{Median projected concentration parameter}& \colhead{} \\
\colhead{Redshift} & \colhead{Connected} & \colhead{Isolated}}

\startdata
     $0.3 \leq z < 0.6$   & $0.70\pm0.07$ &  $0.75\pm0.07$\\
     $0.6 \leq z < 0.9$   & $0.58\pm0.09$ &  $0.58\pm0.08$\\
     $0.9 \leq z < 1.2$   & $0.46\pm0.08$ &  $0.51\pm0.04$\\
      $z \geq 1.2$   & $0.42\pm0.09$ &  $0.45\pm0.09$\\
\enddata 
\end{deluxetable}

\subsection{Comparison with the IllustrisTNG Hydrodynamical Simulation}\label{sec:web_feeding_trend_in_the_TNG_hydronamical_simulation}

To better understand the web feeding model and the related results from the observation in the previous section, we use the IllustrisTNG simulation \citep{springel2018first, nelson2018first}. IllustrisTNG 300-1 (TNG300) has a simulation volume with a box size of \SI{300}{\mega pc} in a side, providing a statistically robust sample of galaxy clusters. The group catalog in IllustrisTNG identifies halos using a standard Friends-of-Friends algorithm \citep{davis1985evolution} with a linking length parameter of $b = 0.2$. Here, $b$ is a dimensionless free parameter that scales the mean inter-particle distance of collapsed halo particles relative to that of the global distance. The commonly adopted value of $b=0.2$ corresponds to a density contrast between halo density and the global mean density to be $200$ \citep{2011ApJS..195....4M}. Our analysis focuses on halos with $M_{200}$ (\texttt{Group\_M\_Crit200}) more massive than $10^{13}$ \SI{}{M_{\odot}} which matches the range of cluster masses observed in COSMOS2020. We also use subhalos derived from the Subfind algorithm which relies on all particle species to identify galaxies \citep{springel2001populating, dolag2009substructures} whose stellar masses within twice the half mass radius are more massive than $10^{8.5}$ \SI{}{M_{\odot}} (see e.g. \citealt{2018MNRAS.473.4077P} for the description of the algorithm). This choice is consistent with a mass complete sample in observation and varying the minimum stellar mass from $10^{8.5}$ to $10^{9}$ \SI{}{M_{\odot}} does not result in different results. We describe details on how the TNG300 data are analyzed to interpret the observational results in Appendix \ref{appendix_a}.
\begin{figure*}
    \centering
    \gridline{\fig{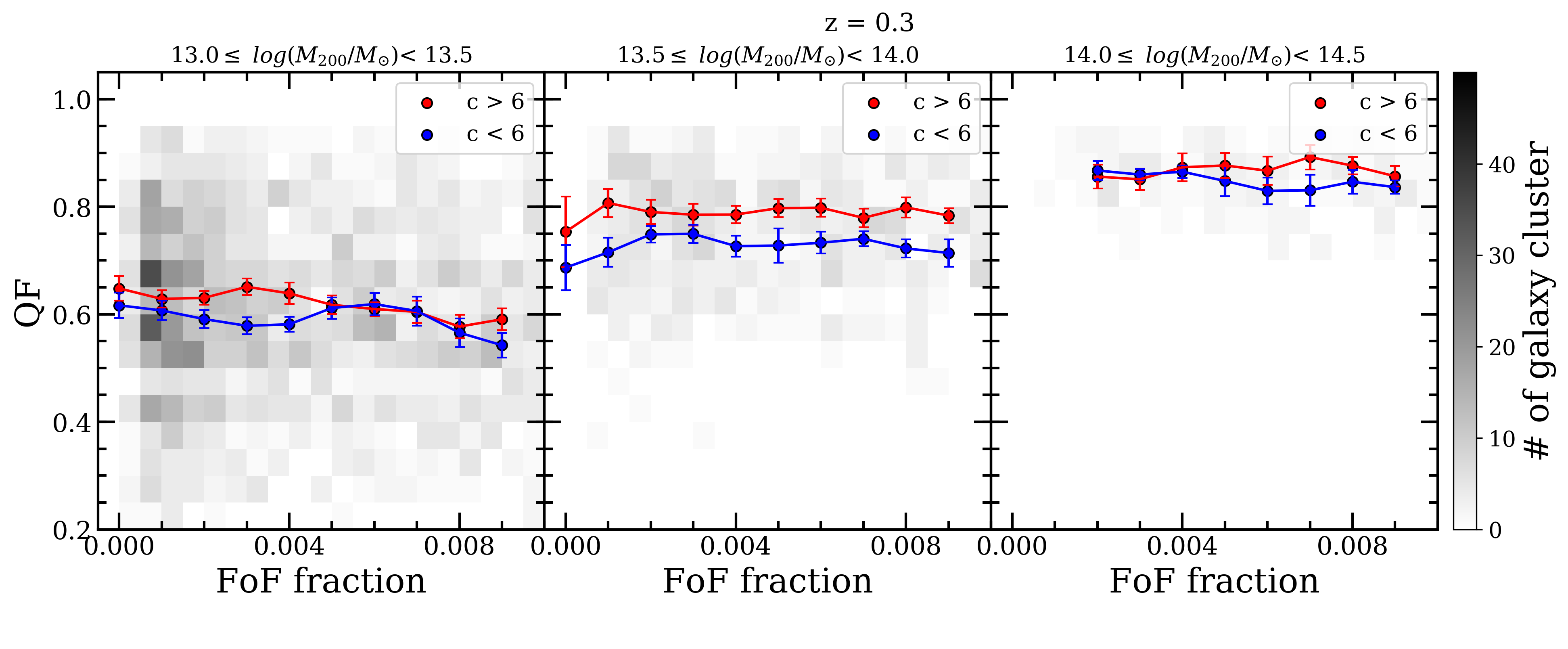}{0.7125\textwidth}{}}
    \gridline{\fig{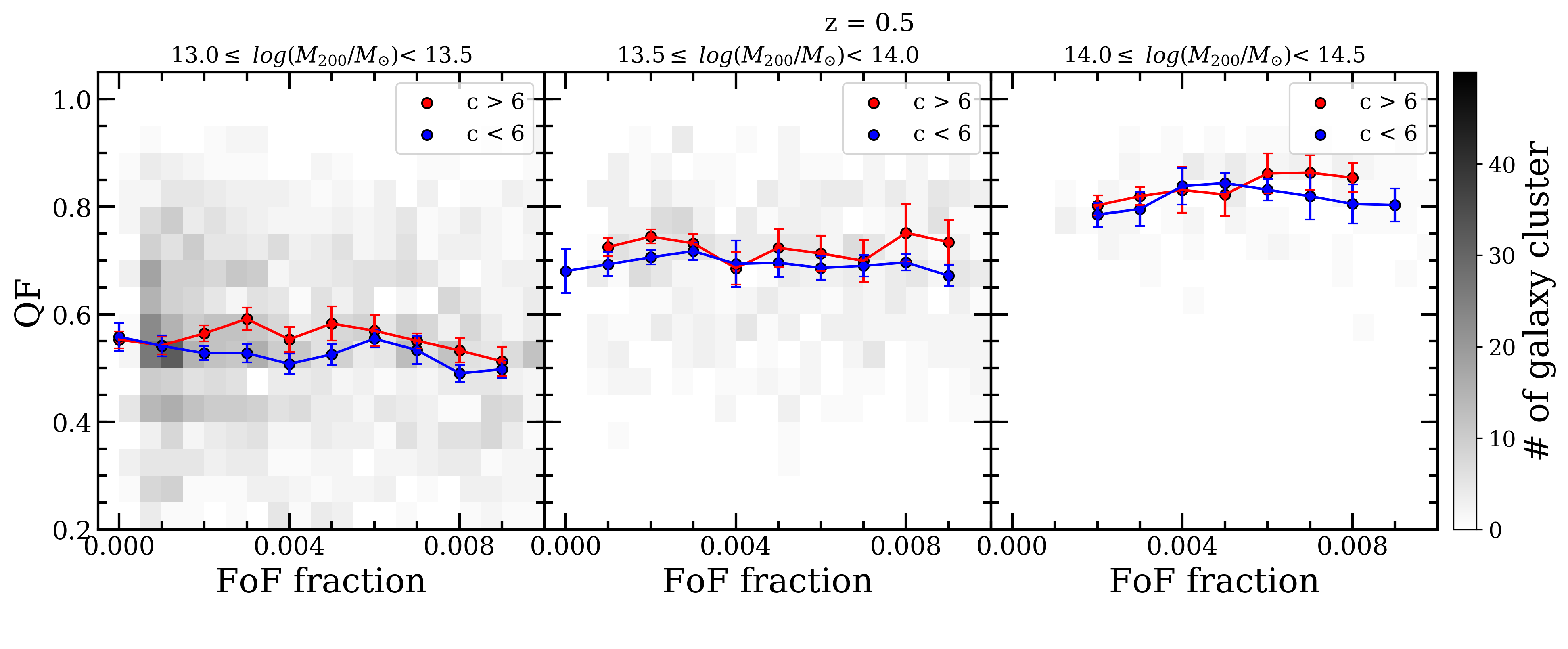}{0.7125\textwidth}{}}
    \gridline{\fig{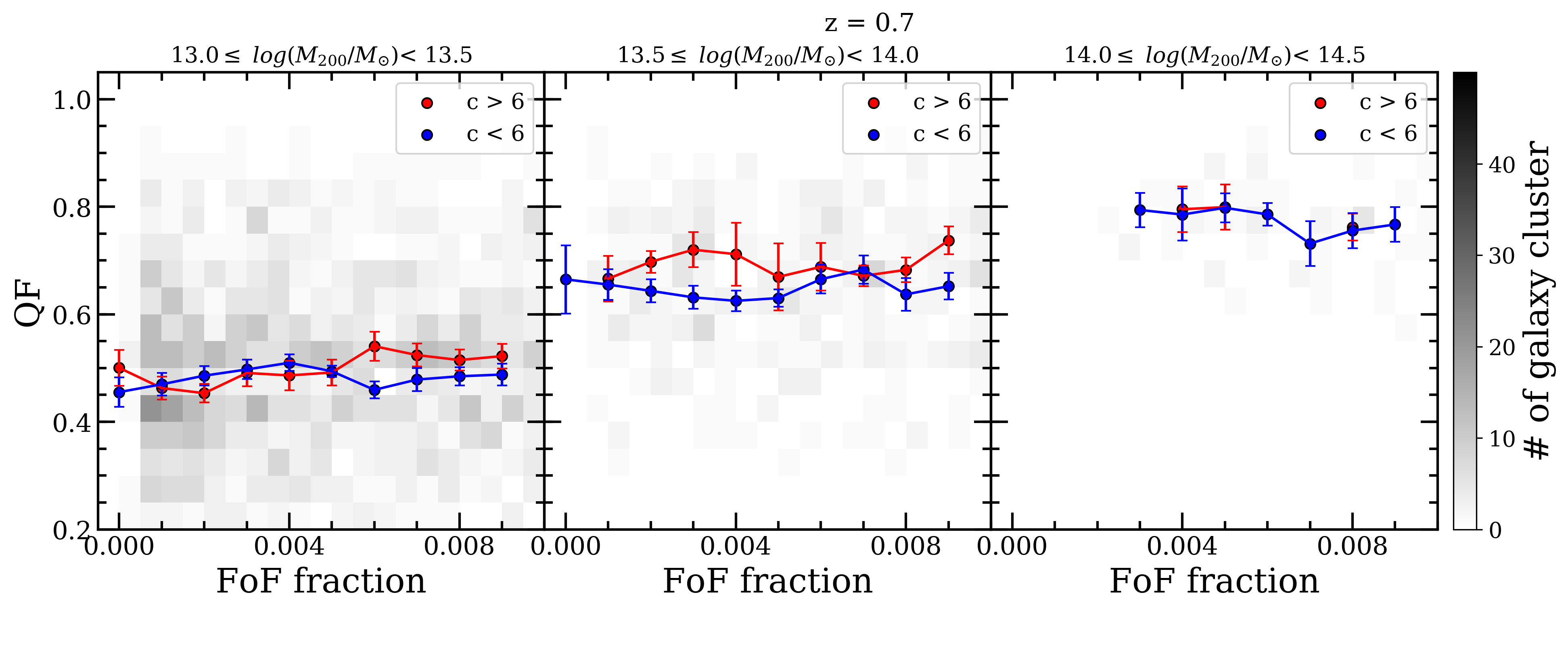}{0.7125\textwidth}{}}
    
    \caption{The median \textit{QF} as a function of \textit{FoF fraction} in the TNG300 simulation. The concentration parameter \textit{c} fitted from the NFW profile \citep{navarro1997universal} is obtained from \citet{anbajagane2022baryonic}.}
    \label{fig:wfm_in_sim}
\end{figure*}
\begin{figure*}
    \centering
    \gridline{\fig{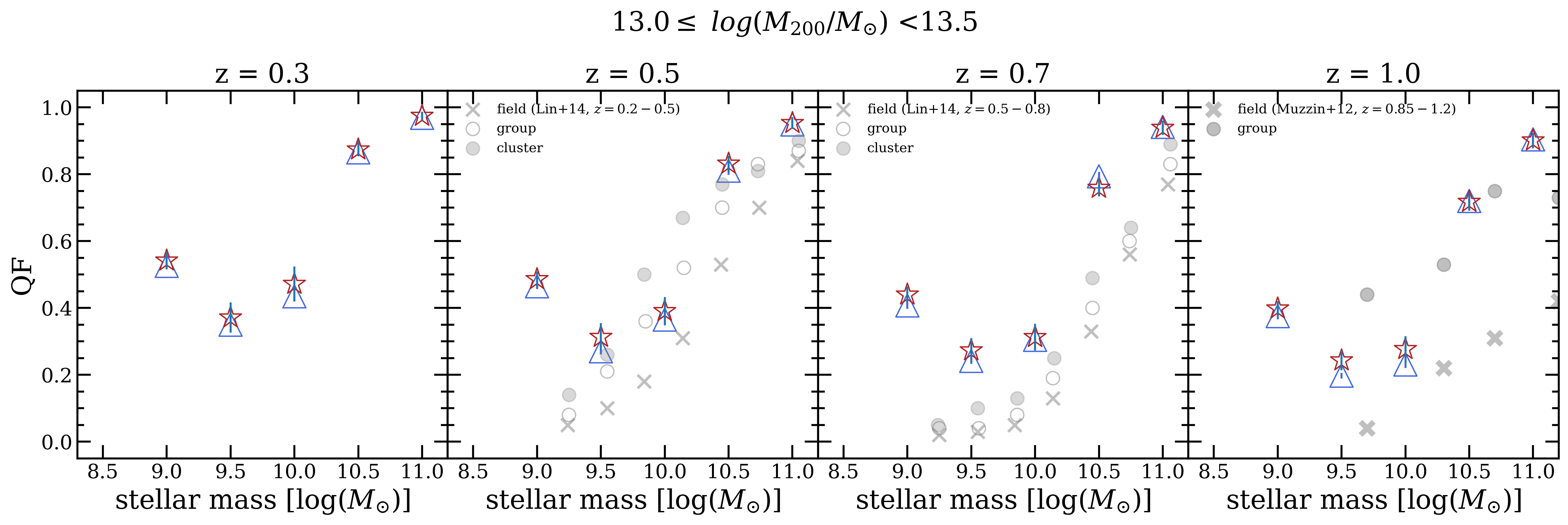}{0.95\textwidth}{}}
    \gridline{\fig{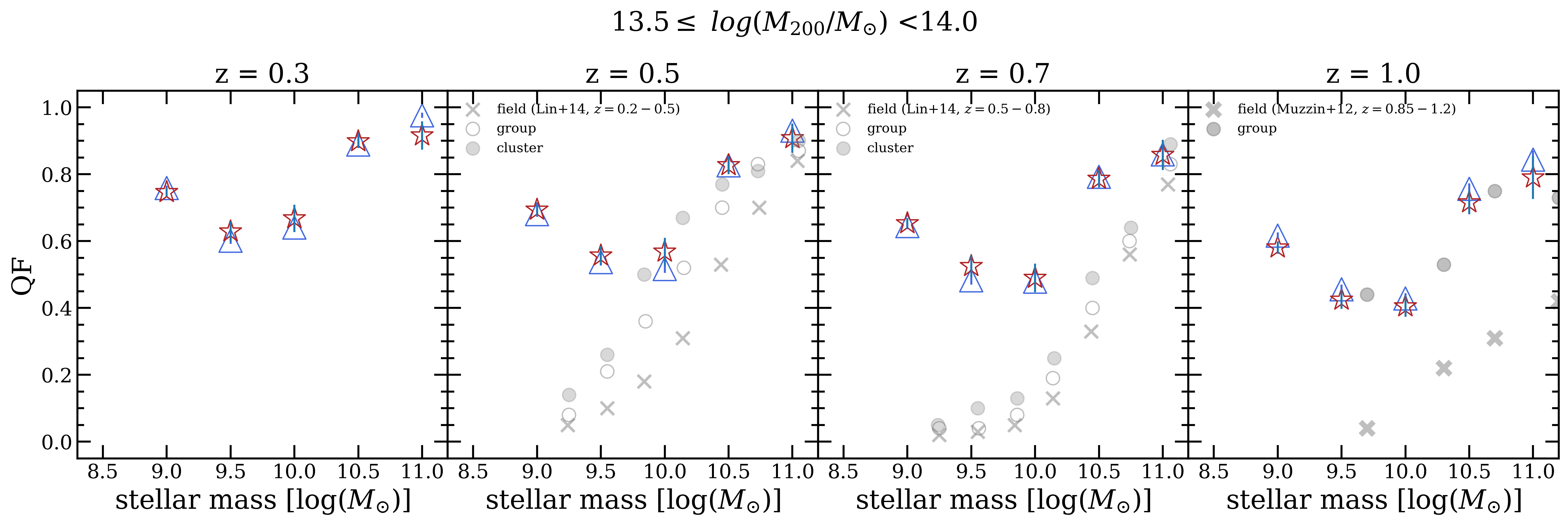}{0.95\textwidth}{}}
    \gridline{\fig{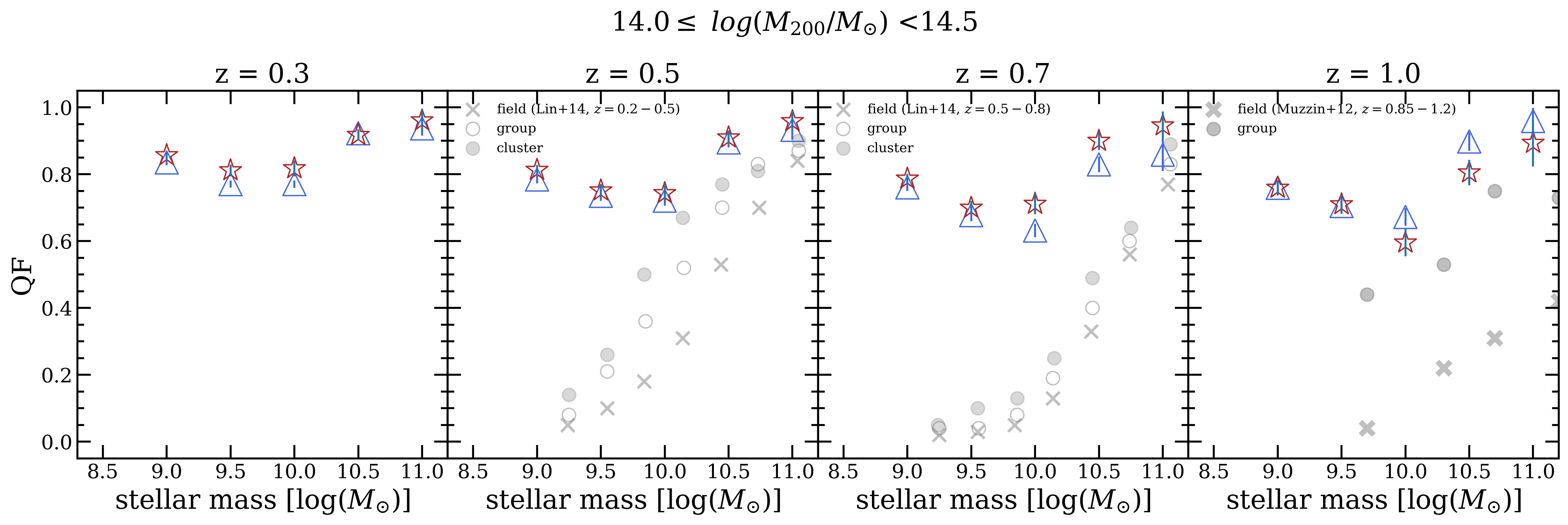}{0.95\textwidth}{}}
    \caption{The median \textit{QF} in stellar mass bins of member galaxies in the TNG300 simulation. The \textit{QFs} of isolated clusters are nearly identical to those in connected clusters over the entire stellar mass bin and regardless of the halo mass. For comparison, the observational data are shown in grey symbols, where the meaning of the points are the same as in Figure \ref{fig:stellarmass_in_obs}.}
    \label{fig:stellarmass_in_sim}
\end{figure*}

We check if the web-feeding trend similar to the result found from COSMOS2020 can be replicated in simulation. The relation between the \textit{FoF fraction} and the median \textit{QF} is shown in Figure \ref{fig:wfm_in_sim}. The \textit{FoF fraction} and \textit{QF} do not seem to be related even after dividing galaxy clusters into different halo masses or high and low-concentration categories. But, \textit{QF} increases with halo mass regardless of redshifts. We compare \textit{QFs} as a function of stellar mass for galaxies in both isolated and connected clusters in Figure \ref{fig:stellarmass_in_sim}. In the simulation results, \textit{QFs} of isolated clusters are nearly identical to those of connected clusters which contradicts the observational results in Figure \ref{fig:stellarmass_in_obs}. An obvious discrepancy between simulation and observation may be found in the distribution of \textit{QF} in stellar mass bins (Figure \ref{fig:stellarmass_in_sim}). In contrast to the increasing trend in \textit{QF} with increasing stellar mass, the \textit{QF} in low stellar mass bins tend to be measured higher. We will speculate on the possible causes for the discrepancy in the next section.

\section{Discussion}
\subsection{Discrepancy between Observations and Simulation}
Considering that the 3D \textit{FoF fraction} is a more accurate representation of the surrounding LSSs than the 2D \textit{FoF fraction}, we expect that the \textit{QF}-\textit{FoF fraction} correlation would be weakened when using the 2D \textit{FoF fraction} with a sizable scatter to trace these structures in comparison to the same relation explored in 3D as in the simulation data. In real, we find the opposite trend as shown in the previous section. Therefore, we exclude the increased scatter in the 2D \textit{FoF fraction} due to moving from the 3D to the 2D distribution from the possible reasons for the discrepancy. Furthermore, we confirm no correlation in TNG300 between \textit{QFs} and \textit{FoF fractions} even when we calculate the 2D \textit{FoF fraction} with the projected 3D \textit{FoF fraction} and repeat the same analysis. 

The tension in the results between the observations and the TNG300 simulation may arise from both observations and simulation. We briefly suggest the possible causes that might drive the disparity.

\subsubsection{Caveats from the Observation}

It is possible that clusters and their member galaxies, determined from photometric redshifts, could suffer from interlopers \citep{2000AJ....120.2851B,2010MNRAS.408.1168B, 2013MNRAS.433.3314S}. While our fiducial choice of photometric redshift uncertainties $\sim 0.01(1+z)$ is obtained from the most up-to-date catalog at the moment, the physical distance corresponding to this error is $\sim$20 Mpc at $z \sim 1$, which is much larger than the typical cluster size. Hence, one may argue that the web-feeding trend could be an outcome of the line-of-sight structures overlapping with each other. We prepared the simulation data as similar as possible to the observational data by adding scatters in redshifts to see if we could reproduce the web-feeding result. However, this test did not produce an artificially created web-feeding effect, suggesting that the line-of-sight effect combined with redshift uncertainty is not likely to solve the tension between observation and simulation.

Finally, the COSMOS2020 field size is smaller than the TNG300 simulation box size, so the cosmic variance may be in play \citep{2010MNRAS.406..881B,2022ARA&A..60..363N}. Such a case can be tested in the future by examining the dataset much wider than COSMOS2020. 

\subsubsection{Caveats from the Simulation}

The resolution limit in large-volume cosmological simulation could be a problem causing the tension. The studies from \citet{donnari2019star}, \citet{donnari2021quenched1}, and \citet{donnari2021quenched2} suggest that \textit{QFs} in simulation can deviate from observations at high mass and low mass end and depending on the halo mass definitions or even \textit{QF} definitions by $10-40\%$. In our case, the \textit{QF} of halos in TNG300 is consistent with the observations at $M_{*} > 10^{9.5}$ \SI{}{M_{\odot}} but at lower masses, it deviates from the observation significantly. We repeat our analysis by restricting the galaxy's stellar mass to $M_{*} > 10^{9.5}$ \SI{}{M_{\odot}} but unfortunately, that does not reveal a correlation between \textit{QF} and \textit{FoF fraction}. 

Matching our cluster samples of interest with simulation is also not trivial. Since our cluster candidates are obtained based on overdensities, we selected samples experiencing various relaxation stages from extended to concentrated structures \citep{2013ApJ...779..127C}. On the contrary, in TNG300, the clusters are detected with the Friends-of-Friends algorithm, finding uniform candidates whose density is 200 times larger than the global one. Due to the difficulties of matching precise definitions of clusters, various quenched fractions are also found even among observations themselves \citep{2013MNRAS.433.3314S,2015MNRAS.452.2528M}.

In short, the web feeding trend observed in COSMOS2020 does not appear in the TNG300 simulation. Future surveys with more accurate redshifts will reduce the uncertainty of cosmic structures and minimize projection effects. The availability of high-resolution cosmological simulation, preferentially the one including a light-cone dataset, will also offer more improved pictures of the effect of large-scale structures on star formation in clusters of galaxies, enabling us to mimic the observed data and analysis in the same way. Given the current limitations, this aspect remains a topic for future analysis. 

\subsection{What Fuels the Galaxy Cluster?}\label{subsec:what_fuels_the_galaxy_cluster}

\begin{figure}
    \centering
    \includegraphics[width = 0.45\textwidth]{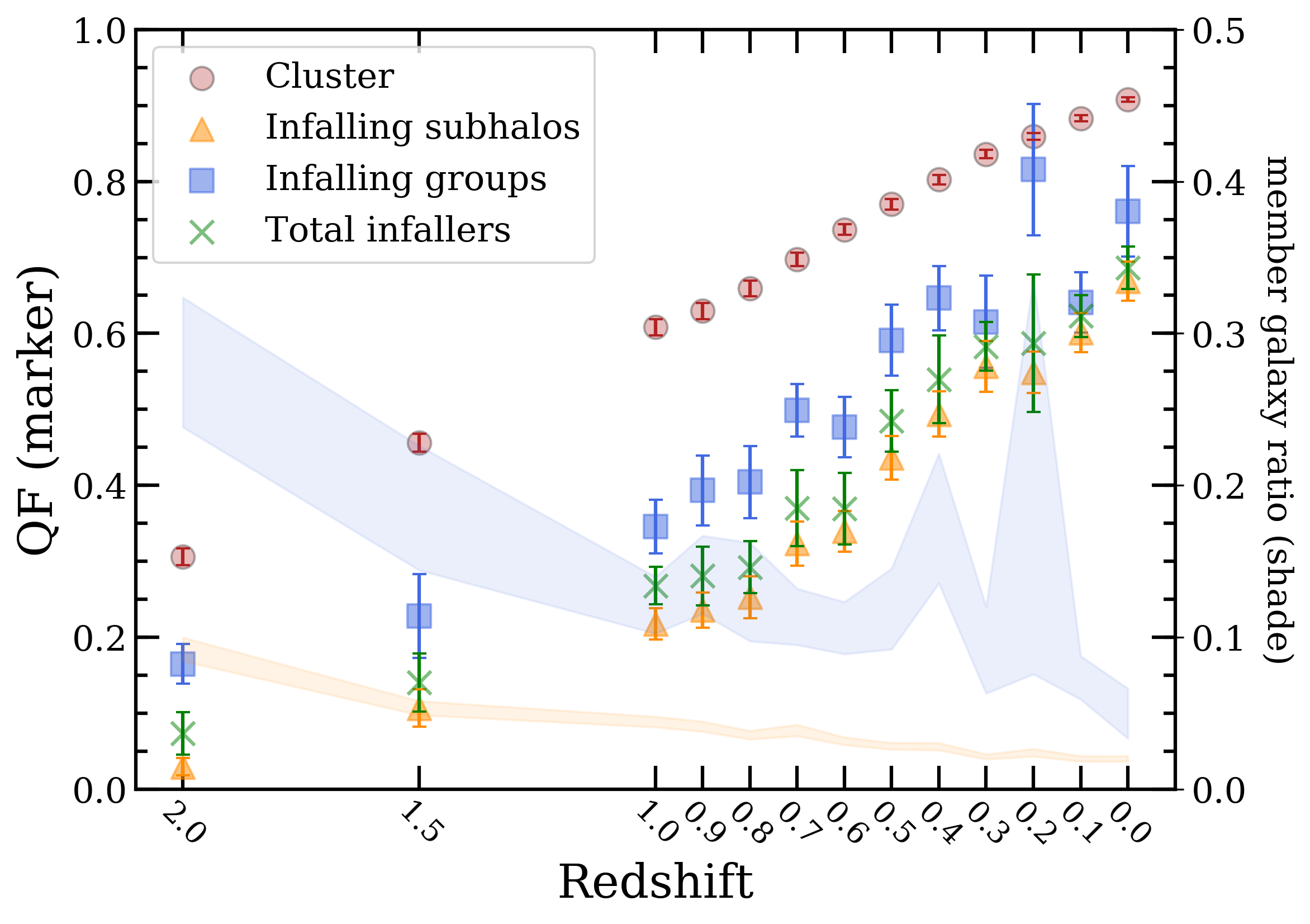}
    \caption{The median \textit{QF} of host galaxy clusters (pink circle), galaxies in infalling groups (blue square), infalling galaxies (orange triangle), and total infallers (galaxies + groups, green cross) at a given snapshot (redshift). The member galaxy ratio, represented as blue (groups) and orange (galaxies) shades, shows the number of infalling galaxies/groups divided by the number of cluster members before the accretion of the galaxies/groups.}
    \label{fig:infalling_qt}
\end{figure}

Here, we focus on the infallers responsible for fueling the host cluster and their respective effects by tracing snapshots at different redshifts in TNG300. By selecting clusters that are more massive than $10^{14}$ \SI{}{M_{\odot}} at the present epoch, we track the member galaxies back in time to $z = 2$. For largely connected galaxy clusters to remain star-forming or less quenched compared to isolated counterparts, \textit{QF} of infalling galaxies must be less than that of the host galaxy cluster. The \textit{QFs} of prospective members, assessed at one snapshot just before their accretion, are depicted in Figure \ref{fig:infalling_qt}. Infalling galaxies embedded in halos more massive than $10^{12}$ \SI{}{M_{\odot}} are classified as infalling groups, and otherwise, individual galaxies. Since not all the nearby galaxies surrounding a given halo infall into the galaxy cluster, we exclusively calculate the \textit{QF} of infallers that would become member galaxies of the host cluster at the next snapshot \citep{donnari2021quenched2, 2022MNRAS.510..581K, 2023MNRAS.518.1316H}. As expected, the \textit{QFs} of both infalling galaxies and groups are lower than the \textit{QF} of the host cluster. For the case of infalling groups, \textit{pre-processing} takes place and the star formation is quenched to a certain degree at the pre-infall stage \citep{1998ApJ...499..589H, 2022ApJS..258...32S}. The infalling individual galaxies have lower \textit{QFs} than galaxies in infalling groups while the number of galaxies in group-scale structures dominates the number of infallers compared to cluster members. The role of infalling groups on cluster evolution is also consistent with previous studies \citep{mcgee2009accretion, donnari2021quenched2}. 

We do not explicitly address the case of cold gas accretion due to the lack of cold gas estimates in COSMOS. Nonetheless, previous studies hint at the role of cold gas accretion in fuelling star formation activities within filaments and cluster environments. We introduce some examples as follows. From xGASS survey \citep{catinella2010galex, catinella2013galex}, \citet{janowiecki2017xgass} shows that central galaxies in low mass groups tend to exhibit higher HI gas fractions and \textit{sSFR} by $0.2-0.3$ dex than galaxies in isolation. They speculate that the HI gas reservoir of low-mass central galaxies is replenished through infalling gas along cosmic filaments and by the merging of gas-rich satellites. Moreover, in regions with moderate overdensities between field and cluster environments, small, gas-rich, and star-forming groups seem to represent an early stage of group evolution. The presence of cold gas and its effect on delayed quenching is also supported by zoom-in cosmological simulations with high resolution. \citet{kotecha2022cosmic} investigate the impact of intra-cluster filaments using hydrodynamic zoom re-simulation of The Three Hundred project \citep{cui2018three, klypin2016multidark}. In the simulation, intra-cluster filaments enable a coherent and less disturbed gas flow, suppressing ram pressure and keeping galaxies forming stars. This shock property of gas can feed clusters through the cosmic web more smoothly \citep{2021MNRAS.502..714R, 2023A&A...673A..62V, 2023A&A...671A.160G, 2024MNRAS.527.1301R}.

\subsection{Can Other Processes Explain the Scatters in Web-Feeding Trend?}

It is important to note that the web-feeding model could be an outcome of various quenching mechanisms acting in galaxy clusters at different evolutionary stages. The scatter in the correlation between \textit{FoF fractions} and \textit{QFs} suggests the involvement of other processes. One possibility is that a delayed quenching timescale might cause the population to diverge from the main correlation. For example, Figure \ref{fig:WFM} shows galaxy clusters with lower \textit{FoF fractions} and lower \textit{QFs}, deviating from the web-feeding trend. We suppose that the delay in quenching might allow isolated clusters to remain star-forming temporarily after web detachment \citep{2013MNRAS.432..336W,2014MNRAS.440.1934T,2015ApJ...806..101H,2018ApJ...866..136F}.

However, long after the web-feeding effect fades away, one may argue that hydrodynamical quenching processes are more prevalent in isolated clusters, where web-feeding is less prominent. With limited gas reservoirs from the cosmic web, the effects of starvation or overconsumption can manifest more dramatically \citep{2014MNRAS.442L.105M, balogh2016evidence}. In dense environments like clusters, gravitational interactions can become more pronounced. As a result, increased chance of mergers \citep{2014ApJ...782...33L, 2017ApJ...845...74J} and tidal stripping \citep{2016MNRAS.463.1907F,2016A&A...596A..11B,2020A&A...638A.133L} could expedite quenching processes. But in such cases, the halo mass should be the main driver for the quenching, rather than the connection to the surrounding LSS.

\section{Conclusion}

We test the web feeding model using the COSMOS2020 data and the IllustrisTNG 300 simulation. Our analysis of the COSMOS field confirms that the large-scale cosmic webs surrounding the galaxy clusters and the star-forming activity are correlated to $z \lesssim 1$. By analyzing the simulation data, we suggest that the correlation possibly results from the infallers supplied by connected overdensities and feeding the galaxy clusters. Our results are summarized as follows.

1. We identify 68 galaxy overdensities from $z = 0.1$ to $1.4$ in the COSMOS field. The halo masses are estimated to be in the range of 13.0 $\leq$ $log(M_{200}/M_{\odot})$ $\leq$ 14.5 by matching them with the X-ray group catalog from \citet{gozaliasl2019chandra}.

2. We find that the quiescent galaxy fraction (\textit{QF}) decreases as redshift increases and halo mass decreases. Nevertheless, there remains a wide range of variation in \textit{QF} of galaxy clusters at a similar redshift and halo mass. The scatters can be at least partially explained by the correlation between \textit{QF} and \textit{FoF fraction} at $z \lesssim 0.9$. For galaxy clusters at $z \lesssim 0.9$, the more connected area (higher \textit{FoF fraction}) shows higher enhancement in star formation activity (lower \textit{QF}), which is consistent with the expectation from the web feeding model. The web feeding model illustrates that the inflow of star-forming galaxies and groups from large-scale structures can keep a galaxy cluster active. 

3. There is no remarkable correlation between \textit{FoF fraction} and \textit{QF} at $z > 0.9$. The \textit{QFs} of cluster members are comparable to those in the field, suggesting that either cluster members have not evolved sufficiently to be distinct from those in the field or that the identification of clusters and cluster members is challenging at the higher redshifts.

4. A complementary perspective is provided by our examination of simulation data. We track the time evolution of galaxy clusters with their surrounding environments using the IllustrisTNG 300 simulation from the present epoch to $z = 2.0$. Unlike in the COSMOS2020, no clear correlation between \textit{QF} and \textit{FoF fraction} can be found. In the simulation, the cause of the discrepancy between the simulation and the observation results is unclear.

5. Using the simulation data, we examine the properties of infalling structures and their paths towards galaxy clusters. Infallers consist of individual galaxies and groups that have lower \textit{QF} than the cluster to which they infall. These infallers follow the FoF overdensities and may contribute to keeping \textit{QF} of clusters low. Group-scale structures encompass the majority of infallers, while individual galaxies contribute to lowering the overall \textit{QF} among infallers.

One limitation of this study is the use of photometric redshifts. Although photometric redshifts are deemed accurate enough for tracing large-scale structures, there is a potential for interlopers to contaminate the measurements of clusters and surrounding large-scale structures. Future studies, supported by a larger number of spectroscopic data, should be able to provide better insights into the connection between cluster star formation activities and surrounding environments.

\begin{acknowledgments}

This work was supported by the National Research Foundation of Korea (NRF) grants, No. 2020R1A2C3011091, and No. 2021M3F7A1084525, funded by the Korea government (MSIT). We would like to thank the COSMOS collaboration. This research was supported in part by grants NSF PHY-1748958 and PHY-2309135 to the Kavli Institute for Theoretical Physics (KITP). S.L. acknowledges support from a National Research Foundation of Korea (NRF) grant (2020R1I1A1A01060310) funded by the Korean government (MIST).

Based on observations collected at the European Southern Observatory under ESO programme ID 179.A-2005 and on data products produced by CALET and the Cambridge Astronomy Survey Unit on behalf of the UltraVISTA consortium.

The IllustrisTNG simulations were undertaken with compute time awarded by the Gauss Centre for Supercomputing (GCS) under GCS Large-Scale Projects GCS-ILLU and GCS-DWAR on the GCS share of the supercomputer Hazel Hen at the High Performance Computing Center Stuttgart (HLRS), as well as on the machines of the Max Planck Computing and Data Facility (MPCDF) in Garching, Germany.

\end{acknowledgments}

\appendix
\counterwithin{figure}{section}
\counterwithin{table}{section}

\section{Mock Simulation with the TNG300} \label{appendix_a}

We provide a detailed description of how we conducted our analysis of the TNG300 simulation to interpret the results obtained from COSMOS2020. The TNG300 identifies galaxy groups with a standard friends-of-friends algorithm run on all kinds of particles (dark matter, gas, stars, black holes) as described in \citet{nelson2018first}. The star formation in TNG300 is implemented by following the procedure of \citet{springel2003cosmological}. Nevertheless, the star formation rate derived in this manner is instantaneous and not compatible with the star formation rate measured in observations. In order to reflect the observational star formation tracer, we adopt a time-averaged star formation rate within appropriate apertures. This adjustment is designed to align the simulation's star formation rate with the observational star formation tracers. Instead of using the star formation rate given in the group catalog directly, we utilize quantities related to the star formation rate from \citet{donnari2019star} and \citet{pillepich2019first}. The star formation rates in the COSMOS2020 catalog are derived by the SED fitting method including IR emission and it reflects the star formation rate in the past $\sim$ \SI{100}{\mega yrs}. Therefore, we employ the time-averaged star formation rate measured over a timescale of the past \SI{100}{\mega yr}. To represent the galaxy-wide star formation activities, the aperture size of twice the stellar half-mass radius is used to calculate the star formation rate. This derived star formation is not completely comparable to this observational study, however, it is known to affect \textit{QF} little because different aperture sizes do not significantly impact the classification between quiescent and star-forming galaxies (see \citealt{donnari2019star} for further details). 

For galaxy clusters, we used the groups with halo masses $M_{200}$ (\texttt{Group\_M\_Crit200}) more massive than $10^{13}$ \SI{}{M_{\odot}} at each snapshot. Groups located within \SI{10}{h^{-1}\mega pc} from the edges of the simulation box are excluded from our analysis. To construct a density field comparable to the observational data, 3-dimensional grid spacing \SI{200}{h^{-1}\kilo pc} is adopted and convolved by a uniform filter of $8\times8\times8$. We note that the number density field derived from COSMOS2020 has a size of \SI{100}{\kilo pc} $\times$ \SI{100}{\kilo pc} $\times$ $0.01(1+z)$ where the redshift uncertainty $\sim 0.01(1+z)$ corresponds to few tens of \SI{}{\mega pc}. In this regard, our choice of grid spaces and convolution scales is designed to contain a similar number of galaxies in each grid cell in the TNG300 simulation. The main difference of \textit{FoF fraction} between the COSMOS2020 and the  IllustrisTNG is that \textit{FoF fraction} is a 3-dimensional cube in place of a 2-dimensional cylindrical volume. We found that the uncertainties derived from the projection effect and photometric redshift do not change the results from the TNG300 when projecting the density field as discussed in Section \ref{sec:web_feeding_trend_in_the_TNG_hydronamical_simulation}. 

\bibliography{references}{}
\bibliographystyle{aasjournal}

\end{CJK*}
\end{document}